\pdfoutput=1
\RequirePackage{etoolbox}
\newbool{fullversion}
\booltrue{fullversion}
\RequirePackage{etoolbox}

\providebool{fullversion}
\providebool{arxivversion}

\documentclass[acmsmall]{acmart}

\usepackage{preamble}
\usepackage{iris}
\usepackage{prob}
\usepackage{examples}

\ifbool{fullversion}{
  \newcommand{\appref}[1]{\cref{#1}}
  \newcommand{\Appref}[1]{\Cref{#1}}
}{
  \newcommand{\appref}[1]{the Appendix~\cite{coneris-full}}
  \newcommand{\Appref}[1]{The Appendix~\cite{coneris-full}}
}

\ifbool{arxivversion}{
\usepackage{caption}
}{
\usepackage[belowskip=-10pt,aboveskip=4pt]{caption}
\setlength{\intextsep}{15pt} %
}

\makeatletter
\ifbool{arxivversion}{}{
\let \MathparLineskip \mpr@lesslineskip %
}
\makeatother

\ifbool{arxivversion}{
  \newcommand{\vsquish}[1]{}
}{
  \newcommand{\vsquish}[1]{\vspace{-#1}}
}

\setcopyright{cc}
\setcctype{by}
\acmDOI{10.1145/3747514}
\acmYear{2025}
\acmJournal{PACMPL}
\acmVolume{9}
\acmNumber{ICFP}
\acmArticle{245}
\acmMonth{8}
\received{2025-02-27}
\received[accepted]{2025-06-27}

\ifbool{fullversion}{
  \citestyle{acmauthoryear}
}{}



\begin{document}

\title{Modular Reasoning about Error Bounds for Concurrent Probabilistic Programs \ifbool{fullversion}{(Extended Version)}{}}

\author[K. H. Li]{Kwing Hei Li}
\orcid{0000-0002-4124-5720}
\affiliation{%
  \institution{Aarhus University}
  \city{Aarhus}
  \country{Denmark}
}
\email{hei.li@cs.au.dk}

\author[A. Aguirre]{Alejandro Aguirre}
\orcid{0000-0001-6746-2734}
\affiliation{%
  \institution{Aarhus University}
  \city{Aarhus}
  \country{Denmark}
}
\email{alejandro@cs.au.dk}

\author[S. O. Gregersen]{Simon Oddershede Gregersen}
\orcid{0000-0001-6045-5232}
\affiliation{%
  \institution{New York University}
  \city{New York}
  \country{USA}
}
\email{s.gregersen@nyu.edu}

\author[P. G. Haselwarter]{Philipp G. Haselwarter}
\orcid{0000-0003-0198-7751}
\affiliation{%
  \institution{Aarhus University}
  \city{Aarhus}
  \country{Denmark}
}
\email{pgh@cs.au.dk}

\author[J. Tassarotti]{Joseph Tassarotti}
\orcid{0000-0001-5692-3347}
\affiliation{%
  \institution{New York University}
  \city{New York}
  \country{USA}
}
\email{jt4767@nyu.edu}

\author[L. Birkedal]{Lars Birkedal}
\orcid{0000-0003-1320-0098}
\affiliation{%
  \institution{Aarhus University}
  \city{Aarhus}
  \country{Denmark}
}
\email{birkedal@cs.au.dk}

\begin{abstract}
  We present \theaplog, the first \emph{higher-order concurrent separation logic} for reasoning
  about error probability bounds of higher-order concurrent probabilistic programs with higher-order
  state. To support modular reasoning about concurrent (non-probabilistic) program modules,
  state-of-the-art program logics internalize the classic notion of linearizability within the logic
  through the concept of \emph{logical atomicity}. In \theaplog, we extend this idea to probabilistic
  concurrent program modules by capturing a novel notion of \emph{randomized logical atomicity} within the
  logic. To do so, \theaplog utilizes \emph{presampling tapes} and a novel \emph{probabilistic
    update modality} to describe how state is changed probabilistically at linearization points. We
  demonstrate this approach by means of smaller synthetic examples and larger case studies.
  All of the presented results, including the meta-theory, have been mechanized in
  the \rocq prover and the Iris separation logic framework.

  \ifbool{fullversion}{This is the extended version of the same paper accepted at ICFP 2025~\cite{coneris-acm}, where more details of proofs and case studies are included in the Appendix.}{}
\end{abstract}

\maketitle

\section{Introduction}
\label{sec:introduction}

Probabilistic data structures, such as
approximate counters, skip lists, or Bloom filters are widely used in concurrent
programming. These data structures can improve time and space efficiency compared to their deterministic counterparts.
However, some probabilistic data structures may return wrong results with a small probability.
Analyzing and ensuring this probability of error is sufficiently small is essential for using these data structures.
But this analysis is challenging because probabilistic programs often have unintuitive behaviors, which are only made more complicated when probabilistic behaviors are combined with concurrency.
Because randomness and concurrency both introduce non-determinism, any analysis must take into account the large range of possible outcomes that can arise from this non-determinism.

For just concurrency alone without randomness, a number of verification techniques have been developed that abstract away from concurrent non-determinism.
For example, Concurrent Separation Logic~(CSL)~\cite{csl} allows for threads in a concurrent program to be verified in a \emph{local} way, without having to consider the effects of interference from other threads at every step of execution. %
A key feature of modern concurrent separation logics is support for proving that data structures are \emph{logically atomic}~\citep{tada,hocap,jacobs-piessens,iris}.
Logical atomicity allows clients to reason \emph{as if} a concurrent data structure had a logical state that is updated atomically at a single point in time during each operation.
This internalizes the idea of \emph{linearizability}, a standard notion of correctness for concurrent data structures, and enables modular and compositional proofs.

It is natural to consider whether similar techniques can be applied to reason about \emph{randomized} concurrent data structures.
Prior work has explored extending concurrent separation logic to the randomized setting~\citep{polaris,cqsl,pcol}.
However, none of these prior logics support compositional reasoning about data structures because they lack support for reasoning about logical atomicity.
Indeed, even developing a suitable notion of what logical atomicity would look like in the presence of randomization is challenging.
Whereas prior efforts developing non-randomized logical atomicitiy could draw inspiration and intuition from the notion of linearizability, there is no widely-accepted analogue of linearizability that is suitable for randomized data structures.
Indeed, prior work has shown ways in which standard notions of linearizability do not suffice when clients of a data structure can make randomized choices~\citep{DBLP:conf/stoc/GolabHW11}. 

In this paper, we develop an appropriate notion of \emph{randomized logical atomicity} in the context of \theaplog, a new concurrent separation logic.
\theaplog is a \emph{probabilistic higher-order concurrent separation logic} for
reasoning about error bounds of \thelang programs, a ML-like language with
support for discrete random sampling, unstructured concurrency, higher-order
functions, and higher-order dynamically-allocated local state.

On the surface level, \theaplog retains all the standard rules of higher-order
separation logic for concurrent programs, along with modern features like expressive impredicative invariants and custom ghost resources.
From the probabilistic side, \theaplog inherits two kinds of separation logic resources from previous logics for reasoning about probabilities, specifically \emph{presampling tapes} first introduced in Clutch~\cite{clutch}, and \emph{error credits} first introduced in Eris~\cite{eris}.
The former are used to reason about random choices that a program will take in the future, while the latter are used to track an upper bound on the probability of some error occurring during execution.

Using its higher-order features, \theaplog encodes randomized logical atomicity by adapting the earlier HOCAP~\citep{hocap} approach to logical atomicity.
To do so, we introduce a novel \emph{probabilistic update modality}, written as $\pupd \prop$.
Intuitively, the modality is used to describe a probabilistic logical update to a piece of ghost state of the program.
In particular, we can use it to reason as if the randomness that the program uses is atomically drawn at one single point in time, which is crucial in writing and proving modular specifications.

We demonstrate the flexibility of our approach by verifying a selection of data structures.
For example, we verify the correctness of a concurrent hash function.
The hash module is subsequently used in an efficient concurrent implementation of a Bloom filter, and we derive a strict error bound for a client program that uses the Bloom filter.
These examples utilize rich language features such as higher-order functions and local state, and display non-trivial interactions between concurrency and probability.
As far as we are aware, even verifying the simpler examples are out-of-scope for previous techniques, let alone reasoning about them \emph{modularly}.
\\

\paragraph{Contributions}
In summary, we make the following contributions:

\begin{itemize}
\item The first concurrent and probabilistic higher-order separation logic for reasoning about error bounds of programs written in \thelang, a probabilistic concurrent higher-order programming language with higher-order references.
\item A novel probabilistic update modality that we use to capture a probabilistic notion of logical atomicity.
\item An extension of the HOCAP approach to the probabilistic setting that allows us to write and prove modular specifications of randomized concurrent data structures.
\item A selection of case studies showcasing our approach to modular verification of higher-order concurrent probabilistic data structures.
\item Full mechanization of all results in the \rocq prover~\cite{coq}, using the Iris separation logic framework~\cite{%
    irisjournal} and the Coquelicot real analysis library~\cite{coquelicot}.
\end{itemize}

\paragraph{Outline} In \cref{sec:tech-overview}, we demonstrate how \theaplog is used to reason about  programs that feature both concurrency and probability, we and highlight some of the key challenges that arise from this combination. We then present the syntax and semantics of our language \thelang in \cref{sec:preliminaries}. In \cref{sec:logic}, we present a collection of program logic rules and we show how to apply them to reason about a concurrent randomized counter module in \cref{sec:modularity}. Subsequently in \cref{sec:case-studies}, we showcase \theaplog on a range of concurrent probabilistic data structure examples\ifbool{fullversion}{}{\footnote{More case studies can be found in the Appendix of the full version of this paper~\cite{coneris-full}}}.  Finally, we discuss related work and conclude with ideas for future work in \cref{sec:related-work} and \cref{sec:conclusion}, respectively.

\section{Motivation and Technical Challenges}
\label{sec:tech-overview}
We first recall the features of the Eris logic~\citep{eris} for reasoning about error bounds of sequential programs, then we discuss how \theaplog{} extends these ideas to the concurrent setting, illustrating some of the challenges that arise when reasoning about error bounds in the presence of concurrency.

\paragraph{Sequential Error Reasoning with Error Credits.}
Eris is a separation logic that introduces a new assertion called \emph{error credits} written $\upto{\err}$, where $0 \leq \err \leq 1$ is a real number.
This assertion represents a budget upper bounding the probability that a specification can fail to hold.
In particular, an Eris Hoare triple of the form $\hoare{P \sep \upto{\err}} \expr {x\Ret. Q}$ implies that if we execute $\expr$ in a state satisfying $P$, then the probability that $\expr$ crashes or returns a value $x$ that violates $Q$ is at most $\err$.

To illustrate how the $\upto{\err}$ assertion is used, consider the following program as a simple example:
\newcommand{\twodie}{\logv{twoAdd}}
\begin{align*}
  \twodie \eqdef{} &\Let l = \Alloc 0 in
  l\gets (\deref l + \Rand 3);
  l\gets (\deref l + \Rand 3);
  \deref l
\end{align*}
Here $\Rand \tapebound$ is a probabilistic construct that samples a value uniformly from $\{0,\dots,\tapebound\}$.
In particular, $\Rand 3$ returns a random number from the set $\{0,\dots,3\}$ with probability $1/4$ each.
The $\twodie$ program first allocates a reference $l$ initialized to $0$, then adds the result of a probabilistic sampling $\Rand 3$ to the value in $l$ twice, and concludes by reading the number in $l$.

Suppose we want to prove an upper bound on the probability that the program returns $0$.
This happens only if both calls to $\Rand 3$ return 0, which occurs with probability at most $1/4 \cdot 1/4 = 1/16$.
We can capture this as a specification in Eris by proving $\hoare{\upto{1/16}}{\twodie}{x\Ret. x > 0}$.

Eris provides 3 key rules for working with error credits:\footnote{We write inference rules with a double horizontal line to mean the rule can be applied in either direction.}
\begin{mathpar}
  {
    \inferHB{err-split}{\upto{\err_1 + \err_2}}{\upto{\err_1} \sep \upto{\err_2}}
  }
  \and
  \inferH{ht-rand-exp}
	{ \expect[\unifd{N}]{\Err} \leq \err }
	{ \hoare{\upto{\err}}{\Rand N}{n \ldotp \upto {\Err(n)} \sep n \in \intrange{0}{N}} }
  \and
  \inferH{err-1}
  {\upto{1}}
  {\FALSE}
\end{mathpar}
The first rule says that error credits can be split and joined together.
The second rule says that when the program makes a randomized choice, we can re-distribute error credits along different branches of the randomized outcome, so long as the \emph{expected value} or \emph{average} amount of error credit does not increase.
Finally, the third rule says that an error credit of $1$ implies $\FALSE$, \ie, we can deduce anything, which intuitively follows from the idea that an upper bound of probability 1 is trivial.
\begin{figure}[t]
  \begin{align*}
    &\{ \upto{\tfrac{1}{16}} \} \\
    & \quad \DumbLet l = \Alloc 0 \In \\
    &\{ l \mapsto 0 \sep \upto{\tfrac{1}{16}} \} \\
    & \quad l\gets (\deref l + \Rand 3); && \text{\footnotesize (apply \ruleref{ht-rand-exp} using  $\Err(x) \eqdef  \tfrac{1}{4} \cdot \iverbr{x = 0}$)} \\
    &\{ \Exists n . l \mapsto n \sep \left((n = 0 \land \upto{\tfrac{1}{4}}) \lor n \neq 0 \right) \} \\
    & \quad l\gets (\deref l + \Rand 3); && \text{\footnotesize (apply \ruleref{ht-rand-exp} using $\Err'(x) \eqdef \iverbr{n = 0 \land x = 0}$)} \\
    &\{ \Exists m . l \mapsto m \sep \left((m = 0 \land \upto{1}) \lor m \neq 0\right) \}  \\
    &\{ \Exists m .  l \mapsto m \sep m \neq 0  \}  && \text{\footnotesize (discharge case $m = 0$ using \ruleref{err-1} from $\upto{1}$)} \\
    & \quad \deref l \\
    &\{ \Ret x . x > 0 \}
  \end{align*}
\caption{Proof outline for the Hoare triple $\hoare{\upto{1/16}}{\twodie}{\Ret x . x > 0}$.}
\label{fig:twodie-seq-proof}
\end{figure}
\cref{fig:twodie-seq-proof} shows a proof outline using these rules to derive the Hoare triple stated above for $\twodie$.
The key steps in this proof are to apply the \ruleref{ht-rand-exp} rule twice to reason about the two $\Rand 3$ statements in the proof.
The first time the rule is applied, we start with $\upto{1/16}$, and instantiate $\Err$ in the rule to be the function $\Err(n) \eqdef \tfrac{1}{4} \cdot \iverbr{n = 0}$ where $\iverbr{P}$ evaluates to $1$ if $P(a)$ is true and to $0$ otherwise.
That is, we end up with $\upto{1/4}$ in the case where $\Rand 3$ returned $0$, and $\upto{0}$ otherwise.
For the latter cases, the remainder of the proof is trivial, since the value in $l$ is already greater than $0$.
For the former case, on the next call to $\Rand 3$, we again apply \ruleref{ht-rand-exp}, setting $\Err'(m) \eqdef \iverbr{n = 0 \land m = 0}$.
Thus, when $\Rand 3$ again returns $0$, we end up with $\upto{1}$.
In this case, $l$ still contains $0$, but we can use \ruleref{err-1} to finish the proof.
For the other cases, $l$ will be greater than $0$, so the postcondition follows directly.

\paragraph{Concurrent Error Bounds in \theaplog}

\theaplog{} generalizes Eris's error credit reasoning to the concurrent setting.
To illustrate some of the key ideas and challenges in doing so, consider the following concurrent variation of the $\twodie$ example:
\begin{align*}
  \contwodie \eqdef{}
  &\Let l = \Alloc 0 in
    \left(\Parallel{\Faa l~(\Rand 3)}{\Faa l~(\Rand 3)}\right);
    \deref l
\end{align*}
Here $|||$ represents parallel composition of two threads, and $\Faa l~n$ is a fetch-and-add command that atomically adds $n$ to the value stored in $l$ and returns the value prior to the addition.
Intuitively, no matter which order the $\Faa$ commands execute in the two threads, the probability that the final $\deref l$ at the end of the program returns $0$ is again bounded above by $1/16$.

\theaplog{} allows us to show this by proving a Hoare triple of a similar form as the one we saw for $\twodie$.
More precisely, in \theaplog, the intuitive meaning of a Hoare triple $\hoare{P \sep \upto{\err}}{e}{Q}$ is that, ``\emph{for all possible schedulings of threads}, if $P$ holds, then the probability that $e$ reaches an error state or returns a value that violates $Q$ is at
most $\err$''.

All of Eris's proof rules for error credits also hold in \theaplog{}.
To reason about parallel execution, \theaplog{} additionally has the familiar parallel composition rule from concurrent separation logic:
\[
  \infrule[right]{ht-par-comp}
          { \hoare{\prop_1}{\expr_1}{\val_1 \ldotp \propB_1\  \val_1 } \\
             \hoare{\prop_2}{\expr_2}{\val_2 \ldotp \propB_2\  \val_2 }
          }
  { \hoare{\prop_1\sep \prop_2}{\Parallel{\expr_1}{\expr_2}}{(\val_1, \val_2) \ldotp \propB_1\  \val_1\sep \propB_2\ \val_2 } }
\]
To apply this rule we have to divide up the precondition into two separate parts, $\prop_1$ and $\prop_2$, and show that they suffice as preconditions for the two threads $\expr_1$ and $\expr_2$, respectively.
We already saw with $\ruleref{err-split}$ that we can split error credits, so for the $\contwodie$ example we might try to split the initial error budget of $\upto{1/16}$ in half, giving each thread $\upto{1/32}$.
However, we would soon run into two issues:
\begin{enumerate}
\item As in the sequential case, we want to apply \ruleref{ht-rand-exp} to try to distribute all of the credits to the cases where the $\Rand 3$ commands return $0$.
  However, if each thread has $\upto{1/32}$ and applies \ruleref{ht-rand-exp} to reason about the $\Rand 3$ it executes, then after applying the rule, it can have at most $\upto{1/8}$ for the case where the $\Rand 3$ returns $0$.
  Combining two $\upto{1/8}$ in the post-condition of the parallel composition rule, we would end up with a total of $\upto{2/8}$ for the case where both threads add $0$ to the counter.
  But this is not enough to apply \ruleref{err-1}, and so we would not be able to prove the intended postcondition.

\item Both threads need to modify the shared location $l$, so they both need to have ``ownership'' of the points-to assertion for $l$. However, $l \mapsto 0 \not\vdash l \mapsto 0 \sep l \mapsto 0$, so we cannot pass ownership of this assertion in the precondition of both threads when applying \ruleref{ht-par-comp}.
\end{enumerate}
The second problem has a well-known solution in CSL, namely \emph{invariant assertions}.
Fortunately, as we will see, it turns out that invariants also provide a solution to the first problem.

\theaplog{} provides Iris-style invariant assertions of the form $\knowInv{}{I}$, which states that an assertion $I$ is an invariant of program execution.
These assertions support the following rules:
\begin{mathpar}
  \infrule[lab]{ht-inv-alloc}
  {\hoare{\knowInv{}{I} \sep \prop}{\expr}{Q}}
  {\hoare{I \sep \prop}{\expr}{Q}}

  \infrule[lab]{inv-dup}
  {\knowInv{}{I} }
  { \knowInv{}{I} \sep \knowInv{}{I}}

  \infrule[lab]{ht-inv-open}
  {\expr\spac \atomic
    \\
    \hoare{I\sep\prop}{\expr}{I\sep\propB}}
  {\hoare{\knowInv{}{I} \sep \prop}{\expr}{\knowInv{}{I} \sep \propB}}
\end{mathpar}
The first rule \ruleref{ht-inv-alloc} says that we can allocate an invariant assertion $\knowInv{}{I}$ if we know that $I$ holds in the precondition.
Invariant assertions are \emph{duplicable}, meaning that we can produce multiple copies using \ruleref{inv-dup}, so that when applying \ruleref{ht-par-comp}, each thread can have a copy of $\knowInv{}{I}$ in its precondition.
Finally, threads can access the assertion $I$ inside of the invariant using \ruleref{ht-inv-open}, which requires that the invariant is restablished after the atomic expression $\expr$ finishes executing.
An expression $\expr$ is  $\atomic$ if it steps to a value in a single execution step.

By putting the $l \mapsto -$ assertion inside of an invariant $I$, we can thus allow both threads to access $l$ by using \ruleref{ht-inv-open} during the $\Faa$ step.
A standard technique in the CSL literature is to use \emph{ghost state} to encode a kind of \emph{protocol} in the invariant assertion that tracks how $l$ can evolve through the shared access by the two threads~\citep{iris}.
We omit the exact details of how this ghost state encoding works, but at a high level, this invariant assertion would have a format like:
\[ I \eqdef \underbrace{\left(l \mapsto 0 \sep \dots\right)}_{\text{no thread added}} \vee \underbrace{\left(\exists v.\, l \mapsto v \sep \dots\right)}_{\text{1 thread added}} \vee \underbrace{\left(\exists v.\, l \mapsto v \sep \dots \right)}_{\text{2 threads added}} \]
where threads use ghost state to track which case of this disjunction they are in.

Our key observation is that if we now also include error credits in the invariant, then we can additionally resolve the first issue mentioned above related to not having a sufficient number of error credits.
For example, by setting the invariant to a form like
\begin{align*}
I \eqdef& \big(\underbrace{\left(l \mapsto 0 \sep \dots\right)}_{\text{no thread added}}
\vee \underbrace{\left(\exists v.\, l \mapsto v \sep \dots\right)}_{\text{1 thread added}}
\vee \underbrace{\left(\exists v.\, l \mapsto v \sep \mhl{v > 0} \sep \dots \right)}_{\text{2 threads added}}\big) \\
& \quad {} \sep
\big(\underbrace{\left(\mhl{\upto{1/16}} \sep \dots\right)}_{\text{no thread sampled}}
\ \vee \underbrace{\left(\mhl{\upto{1/4}} \sep \dots\right)}_{\text{1 thread sampled $0$}}
\ \vee \underbrace{\left(\mhl{\upto{1}} \sep \dots\right)}_{\text{2 threads sampled $0$}}
\ \vee \underbrace{\left(\mhl{\upto{0}} \sep \dots\right)}_{\text{$\geq\! 1$ thread sampled non-$0$}}\big)
\end{align*}
We initially have that the invariant owns the whole error credit $\upto{1/16}$.
The first thread to sample will use this $\upto{1/16}$ with \ruleref{ht-rand-exp}, ending up with $\upto{1/4}$ in the case it samples $0$ and $\upto{0}$ otherwise.
If the first thread samples a non-zero value, then the final value in $l$ will be at least $0$, no matter what the second thread samples.
On the other hand, if the first thread samples $0$, then it will return $\upto{1/4}$ to the invariant, which will then be used by the second thread with \ruleref{ht-rand-exp} to get $\upto{1}$ in the case that it also samples $0$.
At that point, we can use \ruleref{err-1} to exclude this case, just as we did in the original sequential example.
Of course, additional ghost state is needed to track this more complex protocol, but modern CSLs like Iris already provide sophisticated tools for encoding these kinds of protocols.

Although this example seems simple, to the best of our knowledge, \theaplog{} is the \emph{first} unary program logic for randomized concurrent programs that can prove this bound of $1/16$ for $\contwodie$.
Prior concurrent separation logics, even those that are specific to first-order languages~\cite{cqsl, pcol} lack logical facilities necessary for expressing this non-trivial protocol on the shared state.

\paragraph{Modularity and Randomized Logical Atomicity}

While placing error credits in a shared invariant solved the problems discussed in the previous part, the solution we have shown so far is not modular.
To see this issue, imagine that we refactored the code of $\contwodie$ as in \cref{fig:contwodie-refactor}.
\newcommand{\counterpred}{\logv{counter}}
\begin{figure}[t!]
  \centering
  \begin{minipage}[t]{0.4\linewidth}
    \begin{align*}
      \createcounter& \eqdef{} \Lam \_ . \Alloc 0  \qquad \\
      \readcounter& \eqdef{} \Lam l . \deref l \\
      \incrcounter& \eqdef{}  \Lam l. \Faa l\ (\Rand 3)
    \end{align*}
  \end{minipage}
  \begin{minipage}[t]{0.45\linewidth}
    \begin{align*}
      \contwodie \eqdef{}
      &\Let l = \createcounter~\TT in  \\
      &\left(\Parallel{\incrcounter~l}{\incrcounter~l}\right); \\
      &\readcounter~l
    \end{align*}
  \end{minipage}

  \caption{Refactored $\contwodie$ code.}
  \label{fig:contwodie-refactor}
\end{figure}
Here we introduce intermediate functions that encapsulate the operations performed on the location~$l$.
Ideally, we ought to be able to derive separate specifications for these operations, and then use only their specifications to prove the same Hoare triple we had before for $\contwodie$.
For example, we might introduce a predicate of the form $\counterpred\ l\ n \eqdef l \mapsto n$ capturing the value of the counter.
Then, a specification for $\incrcounter$ might look like
\begin{align*}
  \infer
	{ \expect[\unifd{3}]{\Err} \leq \err }
	{\hoare{\counterpred\ l\ x \sep \upto{\err}}{\incrcounter\ l}{\Exists n . \counterpred\ l\ (x + n) \sep \upto{\Err(n)} \sep n \in \intrange{0}{3}}}
\end{align*}
which expresses the effects of adding values to the counter and also allows for averaging error credits across the different outcomes, similarly to \ruleref{ht-rand-exp}.
However, this specification for $\incrcounter$ is not sufficient for verifying $\contwodie$.
As we saw previously for that example, we must put the points-to assertion for $l$ (corresponding to $\counterpred$) and the $\upto{1/16}$ in an invariant.
But we cannot use \ruleref{ht-inv-open} to open this invariant when using the above specification for $\incrcounter$, because $\incrcounter\ l$ is not an atomic expression.

For \emph{non}-probabilistic concurrent programs, a standard solution to this problem is to derive a specification that captures that $\incrcounter$ behaves \emph{as if} it were atomic when incrementing the value in the counter.
Although several different techniques have been proposed for encoding what it means for a function to be \emph{logically} atomic, in \theaplog{} we adapt the HOCAP~\citep{hocap} approach.
At its core, the idea behind HOCAP is to observe that what makes a physically atomic expression special, in terms of the rules of the logic, is that we can open an invariant around it.
Thus, to capture that an operation behaves as if it is atomic, we need a specification style that allows for opening an invariant at the logical point where an operation takes effect.
In Iris, this is captured through the \emph{update modality}, written $\pvs Q$ which holds when $Q$ can be derived by opening invariants.
By deriving a specification for $\incrcounter$ in which the $\counterpred$ assertion occurs under such an update modality in the pre-condition, we can enable $\incrcounter$ to open invariants to get this $\counterpred$ assertion at the moment when it performs the $\Faa$.

To extend this idea to the probabilistic setting, \theaplog{} introduces a new modality, called the \emph{probabilistic update modality} $\pupd$, which additionally allows for error credits to be updated in an expectation preserving way, in the style of \ruleref{ht-rand-exp}.
Intuitively, $\pupd \prop$ holds if we can make an instantaneous probabilistic update of our resources such that the outcome satisfies $\prop$. Compared to the standard update modality $\pvs$, the probabilistic update modality can additionally \emph{redistribute} errors credits through a  logical operation called \emph{tape presampling}  that allows clients to reason about future probabilistic choices.
By using this new modality in HOCAP-style specifications, we are able to capture \emph{randomized} logical atomicity, enabling modular reasoning about concurrent probabilistic libraries.
Because this technique requires more advanced rules of \theaplog{}, we postpone demonstrating it to \cref{sec:tech-overview-modularity}, and first present more of the formal details of \theaplog{} and the semantics of \thelang{}, the concurrent programming language used to express our examples.

\section{Preliminaries}
\label{sec:preliminaries}
In \cref{sec:prelim-prob}, we first  recall various definitions from probability theory. We then present the syntax of \thelang in \cref{sec:the-lang} and its operational semantics in \cref{sec:operational-semantics}.

\subsection{Probability Theory}
\label{sec:prelim-prob}
To account for possibly non-terminating behavior of programs, we define our operational semantics using probability \emph{sub}-distributions. A \defemph{discrete subdistribution} (henceforth simply \defemph{distribution}) on a countable set $A$ is a function $\distr : A \ra [0,1]$ such that $\sum_{a \in A}\distr(a) \leq 1$. The collection of distributions on $A$ is denoted by $\Distr A$. The \defemph{null distribution} \(\nulldistr : \Distr A\) is the constant function \(\lambda x . 0\). We let $\intrange{N}{M}$ denote the set $\{ n \in \nat \mid N \leq n \leq M\}$ and for $N\geq 0$ we let $\unifd{N} \colon \Distr \nat$ denote the (uniform) distribution that returns $1/(N+1)$ for every $n \in \intrange{0}{N}$ and $0$ otherwise.
The \emph{expected value} of $X \colon A \to [0,1]$ with respect to $\distr$ is defined as $\expect[\distr]{X} \eqdef{} \sum_{a \in A} \distr(a) \cdot X(a) $.
The \defemph{mass} of \(\distr\) is $\expect[\distr]{\lambda x. 1}$. Given a predicate $P$ on \(A\), the \emph{Iverson bracket} $\iverbr{P}$ evaluates to $1$ if $P(a)$ is true and to $0$ otherwise,
and the probability of \(P\) \wrt \(\distr\) is \(\pr[\distr] P \eqdef \expect[\mu]{\iverbr{P}}\).
Distributions form a monad; we write $\distr \mbindi f$ for $\mbind(f, \distr)$, which is defined as follows.
\begin{align*}
  &{}\mret \colon A \to \Distr A \qquad &{}\mbind\colon (A \to \Distr B) \times \Distr A \to \Distr B \\
  &{}\mret(a)(a') \eqdef{} \iverbr{a = a'} &{}\mbind(f,\distr)(b) \eqdef{} \sum_{a \in A} \distr(a) \cdot f(a)(b)
\end{align*}

\subsection{The \thelang Language}
\label{sec:the-lang}
Our examples are written in the \thelang language, which is an ML-style programming language extended with probabilistic sampling and fork-based concurrency\footnote{\thelang is also the language studied in Polaris~\cite{polaris}, but we consider probabilistic schedulers in addition to deterministic ones for the full program execution. We discuss this more in \cref{sec:related-work}. }. The syntax of the language is defined by the grammar below:
\begin{align*}
  \val, \valB \in \Val \bnfdef{}
  & z \in \integer \ALT
  b \in \bool \ALT
  \TT \ALT
  \loc \in \Loc \ALT
  \Rec \vf \lvar = \expr \ALT
  (\val,\valB) \ALT
  \Inl \val  \ALT
  \Inr \val \ALT
  \\
  \expr \in \Expr \bnfdef{}  &
  \val \ALT
  \lvar \ALT
  \expr_1~\expr_2 \ALT
  \expr_1 + \expr_2 \ALT
  \expr_1 - \expr_2 \ALT
  \ldots \ALT
  \If \expr then \expr_1 \Else \expr_2 \ALT
  (\expr_1,\expr_2) \ALT
  \Fst \expr \ALT \ldots \\
  & \Alloc~\expr \ALT
  \deref \expr \ALT
  \expr_1 \gets \expr_2 \ALT
  \expr_1 [\expr_2] \ALT
  \Rand \expr \ALT
  \Fork \expr \ALT \Faa \expr_1~\expr_2 %
  \\
  \state \in \State \bnfdef{} & \Loc \fpfn \Val %
  \\
                                \cfg \in \Conf \bnfdef{} & \List (\Expr) \times \State
\end{align*}
The syntax is mostly standard, for example, the expressions $\Alloc \expr$, $\deref\expr$, and $\expr_1 \gets \expr_2$ allocate, load from, and store into a reference, respectively. An array \(\expr_1\) can be accessed at offset \(\expr_2\) (for load or store) via \(\expr_1[\expr_2]\) and  $\Rand \tapebound$ samples from the uniform distribution on $\{0,\dots, \tapebound\}$. 

Concurrency is supported via $\Fork \expr$, which executes \(\expr\) in a new thread, and  \emph{atomic}  operations which provide synchronization between threads.
For example, the atomic fetch-and-add $\Faa\expr_1~\expr_2$ instruction adds the integer $\expr_2$ to the value \(\val\) stored at location $\expr_1$ and returns~\(\val\).\footnote{\thelang also supports other atomic instructions such as atomic exchange and compare-and-swap, which we omit here for brevity.}

A program configuration $\cfg \in \List (\Expr) \times \State$ is given by a pair containing the list of currently executing threads and the heap modeled as a finite map from locations to values.
A configuration \(\cfg\) is \defemph{final} if the first expression in the thread list is a value.

\subsection{Operational Semantics}
\label{sec:operational-semantics}

The operational semantics of \thelang programs is given in stages: expressions take a single execution step, which gets lifted to thread pools by schedulers, and finally these steps are chained together to obtain full program execution.

\paragraph{Expressions} The \(\stepdistr\) function takes an expression (representing the currently active thread) and the current state and produces a distribution over the new expression, new state, and a (possibly empty) list of newly spawned threads.
\thelang has a standard call-by-value semantics where steps can occur under evaluation contexts.
Deterministic language constructs like if-then-else~\eqref{eq:step:if} or \(\Fork \expr\)~\eqref{eq:step:fork} step deterministically by using the return of the distribution monad.
The \(\Rand\tapebound\)  instruction \eqref{eq:step:rand} uniformly associates probability \(1/(\tapebound+1)\) to any integer~\(n\) between 0 and \(\tapebound\).
{\begin{align}
  \stepdistr : (\Expr, \State) \,&\!\to \DDistr {(\Expr,\State, \List(\Expr))} \nonumber\\[1mm]
  \stepdistr (\If \True then \expr_1 \Else \expr_2,\state ) &= \mret (\expr_1, \state, []) \label{eq:step:if} \\
  \stepdistr (\Fork \expr,\state ) &= \mret (\TT, \state, [\expr]) \label{eq:step:fork} \\
  \stepdistr (\Rand \tapebound, \sigma) &= \Lam {(n, \state, [])}\,.\, \textstyle\frac 1 {\tapebound+1} \text{ if } n \in \{ 0, \ldots, N \} \text{\quad and\; \(0\)\; otherwise} \label{eq:step:rand}
\end{align}}

\paragraph{Thread Pools and Schedulers}
The operational semantics of a configuration \(\cfg = (\vec \expr, \state)\) is then given simply by indicating which thread amongst \(\vec \expr\) should step, \ie, by specifying an index \(\threadid \in [0, |\vec \expr|-1]\) and applying the step function to \((\expr_\threadid, \state)\):
\[
  \tpstepdistr(\vec \expr, \state)(\threadid) \eqdef
  \begin{cases}
    \nulldistr & \text{if \((\vec \expr, \state)\) is final,}\\
    \mret (\vec \expr, \state) & \text{if \(\expr_\threadid\) is a value,}\\
    \stepdistr(\expr_\threadid, \state) \mbindi
    \Lam(\expr_\threadid', \state', \vec \expr').
    \mret (\vec \expr [\threadid \mapsto \expr_\threadid'] \dplus \vec \expr' , \state') & \text{otherwise.}
  \end{cases}
\]
If \(\cfg\) is final, it does not step. If \(\expr_\threadid\) is a value, we take a stutter step. Otherwise, we update the \(\threadid\)-th thread with the stepped expression~\(\expr_\threadid'\) and append the newly spawned threads \(\vec \expr'\) to the thread pool.

A \defemph{scheduler} decides which thread in a configuration to step next. Formally, a (probabilistic, stateful) scheduler is given by a transition function $\sch : (\Schstate \times \Cfg) \to \Distr{\Schstate \times \nat}$, which takes in an internal state \(\schstate \in \Schstate\) and a configuration~\(\cfg\), and returns a distribution on its new internal state and the index of the thread in~\(\cfg\) to step next.

Given a configuration $\cfg$, a scheduler $\sch$, and a scheduler state $\schstate$, we can now define the single scheduler-step reduction function $\schstepdistr{\sch}{\schstate,\cfg} \in \Distr{\Schstate\times\Cfg}$ as follows:
\begin{align*}
  \schstepdistr{\sch}{\schstate,\cfg} \;\eqdef\quad
  \sch (\schstate, \cfg)
  \mbindi  \Lam (\schstate', \threadid) . \tpstepdistr(\cfg, \threadid)
  \mbindi \Lam \cfg' . \mret (\schstate', \cfg')
\end{align*}
Intuitively, $\schstepdistr{\sch}{\schstate,\cfg}$ first evaluates $\sch (\schstate, \cfg)$ to get a new state $\schstate'$ and index~$\threadid$, steps the \(\threadid\)-th thread to obtain the configuration \(\cfg'\), and returns the new scheduler state and configuration.

The notion of scheduler we consider is quite strong. Firstly, the scheduler is \emph{probabilistic} and can update its internal state and choose the next thread to step probabilistically instead of deterministically. Secondly, the update decision of a scheduler can depend not only on its internal state, but also on the \emph{entire} view of the thread pool and the memory state. These two design choices provide more power to the scheduler and enable us to reason about the error bounds of algorithms under a larger and richer class of schedulers than, say, deterministic schedulers.

\paragraph{Program Execution}
We next define \(n\)-step program execution with respect to a scheduler~\(\sch\) as the following recursive function  $\execVal_{\sch, n}:(\Schstate\times\Cfg) \to \Distr\Val $.
\[ \execVal_{\sch, n}(\schstate, \cfg) \eqdef
  \begin{cases}
    \mret\val
    & \text{if \(\cfg\) is final and \(\cfg = (\val \cons \vec e, \state)\) for some \(\val \in \Val\)}, \\
    \nulldistr
    &  \text{if \(n=0\) and } \cfg \text{ is not final}, \\
    \schstepdistr{\sch}{ \schstate, \cfg} \mbindi \execVal_{\sch, n-1}
    & \text{otherwise.}
  \end{cases} \]
One can read $\execVal_{\sch, n}(\schstate, \cfg)(\val)$ as the probability of returning $\val$ in the first thread after at most~$n$ steps of~$\cfg$ under the scheduler~$\sch$ initialized with the scheduler state~$\schstate$.
Finally, full program execution is defined as the limit of \(\execVal_{\sch, n}\), which exists by monotonicity and continuity:
\[  \limexecVal_{\sch}(\schstate, \cfg) \eqdef \lim_{n\rightarrow \infty} \execVal_{\sch, n} (\schstate, \cfg) \]
We simply write $\limexecVal_{\sch}\, \expr$  if the result is the same for all initial program and scheduler states.

Traditionally, a program \(\cfg\) is \emph{safe} if it never gets stuck during execution, \ie, any partial program execution starting from \(\cfg\) is either a value or it can make progress.
To define the appropriate notion of safety for the probabilistic setting of \thelang (see \cref{thm:safety}), we need the following auxiliary definition of partial program execution $\pexec_{\sch,n}: (\Schstate\times\Cfg)\rightarrow\DDistr(\Schstate\times\Cfg)$.
\begin{align*}
  \pexec_{\sch, n}(\schstate, \cfg) &=
                                      \begin{cases}
                                        \mret(\schstate,\cfg) & \text{if \(\cfg\) is final or \(n=0\)}, \\
                                        \schstepdistr{\sch}{ \schstate, \cfg} \mbindi \pexec_{\sch, n-1} & \text{otherwise.}
                                      \end{cases}
\end{align*}
We can view \(\pexec\) as a relaxation of \(\execVal\) which keeps probability mass on configurations that are not final, whereas the latter only considers final configurations. %

\section{Logic}
\label{sec:logic}

In this section, we dive into the rules of \theaplog.
We start with a glance of the syntax of the logic and its adequacy theorem.
Then, we explore the general program logic rules before discussing presampling tapes and the probabilistic update modality.

\subsection{Introduction to \theaplog}
\label{subsec:intro-logic}
The \theaplog logic is built on top of the Iris base logic~\cite{irisjournal} and inherits all of the basic propositions and their associated proof rules.
This includes the \emph{later} modality $\later$, the \emph{persistence} modality $\always$ and the points-to connective $\loc \mapsto v$ that asserts exclusive ownership of the location $\loc$ storing value $\val$.
A selection of \theaplog propositions are shown below.
\begin{align*}
  \prop,\propB \in \iProp \bnfdef{}
  & \TRUE \ALT \FALSE \ALT \prop \land \propB \ALT \prop \lor \propB \ALT \prop \Ra \propB \ALT
    \All \var . \prop \ALT \Exists \var . \prop \ALT \prop \sep \propB \ALT \prop \wand \propB \ALT \later \prop \ALT \always \prop \ALT \\
  & \knowInv{\iname}{\prop} \ALT
    \ownGhost{\gname}{\ghostRes} \ALT
    \pvs[\mask_{1}][\mask_{2}] \prop \ALT
    \progheap{\loc}{\val} \ALT
    \hoare{\prop}{e}{\propB}[\mask] \ALT
    \upto{\err} \ALT
    \progtape{\lbl}{\tapebound}{\tape} \ALT
    \pupd[\mask_{1}][\mask_2]\prop\ALT
    \ldots
\end{align*}
\theaplog is a separation logic and propositions denote sets of resources.
$\prop \sep \propB$ holds for resources that can be decomposed into two disjoint pieces satisfying $\prop$ and $\propB$.
The separating implication $P \wand Q$ is the right adjoint of $\sep$, in the sense that $P \sep (P \wand Q) \vdash Q$.
While omitted in \cref{{sec:tech-overview}}, note that invariant assertions $\knowInv{\iname}{P}$ are annotated with an identifying name $\iname$ which is used to prevent the prover from opening the same invariant twice (which is unsound).
For bookkeeping purposes, Hoare triples are annotated with the set of invariant names that the specification relies on; we omit this \emph{mask} annotation when considering the set of all invariant names $\top$.

As mentioned in \cref{sec:tech-overview}, we internalize error bounds using the error credit assertion $\upto{\err}$.
The \emph{presampling tape} assertion $\progtape{\lbl}{\tapebound}{\tape}$ is a probabilistic connective that we adapt from Clutch \cite{clutch} which  plays a key role in deriving certain modular specifications.
The probabilistic update modality $\pupd[\mask_1][\mask_2]\prop$ is a novelty of the \theaplog logic.
We further discuss these three connectives and their role in the following section.

The meaning of the \theaplog Hoare triple is captured by the adequacy theorem shown below.
\begin{theorem}[Adequacy]\label{thm:adequacy}
  If $\hoare{\upto \err}{\expr}{\pprop}$,
  then for all schedulers $\sch$,
  $\pr[\limexecVal_\sch \expr]{\neg \pprop} \leq \err$. %
\end{theorem}

The theorem says that by proving a Hoare triple for the expression $\expr$, assuming initial ownership of $\upto{\err}$ error credits, then for all schedulers $\sch$, the probability of the program $\expr$ returning a value\emph{ not} satisfying the proposition $\pprop$ is smaller than or equal to $\err$.

In addition, we have another safety theorem that provides an upper bound on the probability of the expression getting stuck.
\begin{theorem}[Safety]\label{thm:safety}
  If $\hoare{\upto \err}{\expr}{\TRUE}$,
  then for all schedulers $\sch$ with mass $1$ and integers $n$,
  the mass of $\pexec_{\sch, n} {(\schstate,([\expr],\state))}$ is greater or equal to $1-\err$. %
\end{theorem}
Intuitively, this theorem states that proving $\hoare{\upto \err}{\expr}{\TRUE}$ in \theaplog implies that the probability of $\expr$ getting stuck is at most~$\err$ for all schedulers\footnote{The condition that \(\sch\) must have mass~1 (for all scheduler states) rules out the pathological situation where a configuration is ``stuck'' because \(\sch\) has probability less than 1 to pick \emph{any} thread to step next.
  The assumption is only used in \cref{thm:safety}.}.

\subsection{Rules of \theaplog}
\label{subsec:rules}

\paragraph{Program-Logic Rules}
Coneris satisfies all the usual structural and computational rules present in Iris-based separation logics.
For example, \theaplog satisfies the bind rule (\ruleref{ht-bind}), the frame rule (\ruleref{ht-frame}), and the usual computational rules for interacting with the heap (\eg{}, \ruleref{ht-load}).
\begin{mathpar}
  \inferH{ht-bind}
  { \hoare{\prop}{\expr}{\val . \propB} \\  \All \val. \hoare{\propB}{\fillctx\lctx[\val]}{\propC}}
  { \hoare{\prop}{\fillctx\lctx[\expr]}{\propC}}
  \and
  \inferH{ht-frame}
  {\hoare{\prop}{\expr}{\propB}}
  { \hoare{\prop\sep\propC}{\expr}{\propB\sep\propC}}
  \and
  \inferH{ht-load}
  {}
  {\hoare{l\mapsto v}{\deref l}{w.w=v\sep l\mapsto v}}
\end{mathpar}

Invariants can be allocated by giving up ownership of the corresponding resources (\ruleref{ht-inv-alloc}).
If we own an invariant, we can temporarily, for one atomic step, get access to its contents (\ruleref{ht-inv-open}).
The later modality $\later$ is important for soundness but can otherwise be ignored \cite{irisjournal}.
\begin{mathpar}
  \inferH{ht-inv-alloc}
  { %
    \hoare{\knowInv{\iname}{\prop} \sep \propB}{\expr}{\propC}[\mask] }
  { \hoare{\prop \sep \propB}{\expr}{\propC}[\mask] }
  \and
  \inferH{ht-inv-open}
  {\expr\spac \atomic \\
    \hoare{\later I\sep\prop}{\expr}{\later I\sep\propB}[\mask]}
  { \hoare{\knowInv{\iname}{I} \sep \prop}{\expr}{\propB}[\mask\uplus\{\iname\}]} \and
\end{mathpar}

\paragraph{The Update Modality}
The \emph{update modality} $\pvs[\mask_{1}][\mask_{2}]$ is the primary primitive for manipulating ghost resources and interacting with invariants in the Iris base logic.
As we alluded to earlier in \cref{sec:tech-overview}, a key idea behind the HOCAP approach to modular specification is to use this modality as a way to assert that a proposition could be proven by opening invariants.

The update modality $\pvs[\mask_{1}][\mask_{2}]$ is annotated with two sets of invariants.
We write $\pvs[\mask]$ when $\mask_{1} = \mask_{2} = \mask$ and $\pvs$ when $\mask = \top$, the set of all names.
Intuitively, the assertion $\pvs[\mask_{1}][\mask_{2}] \prop $ denotes a resource that, together with the resources from the invariants in $\mask_1$, can be updated and split into two disjoint pieces: one satisfying $\prop$ and one satisfying the invariants in $\mask_{2}$.
That is, we can use the update modality to \emph{specify} resource updates and invariant access (\ruleref{inv-open}) as an \emph{assertion} in the logic rather than just as a primitive rule of the program logic.
The update modality can be eliminated (\ruleref{ht-fupd-elim}) at any suitable time during program verification.
\begin{mathpar}
  \inferH{inv-open}
  {\knowInv{\iname}{\prop} \\
    \later \prop \wand \pvs[\mask_1][\mask_2] (\later \prop \sep \propB) }
  {\pvs[\mask_1\uplus\{\iname\}][\mask_2\uplus\{\iname\}] \prop}
  \and
  \inferH{ht-fupd-elim}
  {\expr\spac \atomic \\
  \hoare{\prop \sep \propB}{\expr}{\pvs[\mask_2][\mask_1] \propC}[\mask_2]}
  {\hoare{\big(\pvs[\mask_1][\mask_2] \prop \big) \sep \propB}{\expr}{\propC}[\mask_1]}
\end{mathpar}
A key idea behind the approach we apply in \cref{sec:modularity} is to parameterize program specifications by a proposition of the shape $\prop \wand \pvs[\mask_{1}][\mask_{2}] \propB$, a so-called \emph{view shift}, that is eliminated at the linearization point of the module operation.
By providing a view shift as an argument, the \emph{client} can specify how they wish for their logical state (their ``view'') to evolve when the operation physically takes place.

\paragraph{Presampling Tapes}
Reminiscent of how prophecy variables \cite{iris-prophecy, prophecy1, prophecy2} allow us to talk about the future, presampling tapes give us the means to talk about the outcome of sampling statement in the future.
Presampling tapes were introduced in Clutch \cite{clutch} to address an alignment issue in refinement proofs, but as we later see in \cref{sec:modularity} they also play a crucial role in modularizing (unary) proofs about concurrent probabilistic programs through probabilistic view shifts.

Intuitively, presampling tapes allow us \emph{in the logic} to presample the outcome of future sampling statements.
Formally, they appear both operationally and in the logic.
In the programming language, presampling tapes appear as two new ghost code constructs, $\AllocTape \expr$ and $\Rand \expr_1\ \expr_2$, that are used to allocate a new presampling tape and sample from a tape, respectively.

\begin{minipage}{0.5\linewidth}
  \begin{align*}
  \val \in \Val \bnfdef{}& \ldots \ALT \lbl \in \Lbl \\
  \expr \in \Expr \bnfdef{}& \ldots \ALT
                             \AllocTape\,\expr \ALT
                             \Rand \expr_{1}~\expr_{2}
  \end{align*}
  \strut
\end{minipage}
\hfill
\begin{minipage}{0.5\linewidth}
  \begin{align*}
  \state \in \State \eqdef{}& (\Loc \fpfn \Val) \times (\Lbl \fpfn \Tape) \\
    t \in \Tape \eqdef{}& \{ (\tapebound, \tape) \mid \tapebound \in \mathbb{N} \wedge \tape \in \mathbb{N}_{\leq \tapebound}^{\ast} \}
  \end{align*}
  \strut
\end{minipage}

In the operational semantics, allocation of a fresh presampling tape~\eqref{eq:step:alloctape} via \(\AllocTape \tapebound\) deterministically associates a fresh label~\(\lbl\) to the empty tape~\(\nil\).
A labelled \(\Rand \lbl\ \tapebound\) with an empty tape samples uniformly~\eqref{eq:step:randnil}, \ie, it behaves like an unlabelled \(\Rand \tapebound\).
If, on the other hand, the tape \(\lbl\) is non-empty, \(\Rand \tapebound\ \lbl\) deterministically pops the first value~\(n\) from the tape~\eqref{eq:step:randcons}.
Note that no step in the operational semantics \emph{writes} contents to a presampling tape.
In fact, tapes and label annotations do not in any way alter the behavior of the program  and can be entirely erased \cite{clutch}.
However, as we later see, the probabilistic update modality  allows us to reason \emph{as if} a presampling step could asynchronously pre-populate a tape with a random sample at any point in time.
{\small\begin{align}
  \stepdistr(\AllocTape \tapebound, \state) &= \mret (\lbl,\lupdate{\state}{\lbl}{(\tapebound,\nil)} , []) \quad \text{(where $\lbl$ is fresh \wrt \( \state\))} \label{eq:step:alloctape}\\
  \stepdistr (\Rand \lbl\ \tapebound, \sigma) &= \Lam {(n, \state, [])}\,.\, \textstyle\frac 1 {\tapebound+1} \text{ if } \state[\lbl] = (\tapebound, \nil) \land n \in \{ 0, \ldots, N \} \text{\quad and\; \;} 0 \text{ otherwise} \label{eq:step:randnil} \\
  \stepdistr(\Rand \lbl\ \tapebound, \lupdate{\sigma}{\lbl}{n\lapp\tape}) &= \mret (n,\lupdate{\sigma}{\lbl}{\tape} , []) \label{eq:step:randcons}
\end{align}}

Now, the logical assertion $\progtape{\lbl}{\tapebound}{\tape}$ denotes ownership of the presampling tape $\lbl$ with bound $\tapebound$ and contents $\tape$, analogously to how the points-to connective for the heap denotes ownership of a location and its contents.
The two rules \ruleref{ht-alloc-tape} and \ruleref{ht-rand-tape} reflects the operational behavior of equations~\eqref{eq:step:alloctape} and~\eqref{eq:step:randcons} in the logic.
\begin{mathpar}
  \inferH{ht-alloc-tape}
  {}
  {\hoare{\TRUE}{\AllocTape~\tapebound}{ \lbl \ldotp \progtape{\lbl}{\tapebound}{\nil}}}
  \and
  \inferH{ht-rand-tape}
  { }
  { \hoare
    {\progtape{\lbl}{\tapebound}{n\cdot\tape}}
    {\Rand \lbl~\tapebound}
    {\Ret x .  x = n \sep \progtape{\lbl}{\tapebound}{\tape}} }
\end{mathpar}

\paragraph{Probabilistic Update Modality}
Previous work \cite{clutch, eris, approxis} introduce presampling tapes for different purposes but, common for all instantiations, presampling is only supported as a rule in the program logic.
Similar to how \ruleref{ht-inv-open} only allows us to reason about invariants using the program logic, presampling is only supported as a primitive program-logic rule.
This is \emph{not} sufficient for the modular specifications we set out to prove.
Intuitively, we need a way to specify updates to presampling tapes as an \emph{assertion}, just like the update modality enables us to specify invariant access and resource updates as an assertion.

To this end, we introduce the \emph{probabilistic update modality} $\pupd[\mask_1][\mask_2]\prop$.
This modality satisfies all the same rules as the update modality: \eg{}, it can be used to open invariants (hence the invariant masks $\mask_{1}$ and $\mask_{2}$) and update resources, it is monadic (\ruleref{pupd-ret} and \ruleref{pupd-bind}), can be derived from the update modality (\ruleref{pupd-fupd}), and it can be eliminated (\ruleref{ht-pupd-elim}).
\begin{mathpar}
  \inferH{pupd-ret}
  { \prop
  }
  {\pupd[\mask] \prop}
  \and
  \inferH{pupd-bind}
  { \pupd[\mask_1][\mask_2] \prop \\
    \prop \wand \pupd[\mask_2][\mask_3] \propB
  }
  {\pupd[\mask_1][\mask_3] \propB}
  \and
  \inferH{pupd-fupd}
  {\pvs[\mask_1][\mask_2] \prop
  }
  {\pupd[\mask_1][\mask_2] \prop}
  \and
  \inferH{ht-pupd-elim}
  {\hoare{\prop \sep \propB}{\expr}{\propC}[\mask]
  }
  {\hoare{(\pupd[\mask] \prop) \sep \propB}{\expr}{\propC}[\mask]}
  \and
  \inferH{pupd-presample-exp}
  {
    \expect[\unifd{N}]{\Err} \leq \err \\
    \upto{\err} \\
    \progtape{\lbl}{\tapebound}{\tape}
  }
  {\pupd[\mask]
    (\Exists n. \progtape{\lbl}{\tapebound}{\tape\cons n} \sep \upto{\Err(n)} \sep n \in \intrange{0}{N})}
  \and
  \infrule[lab]{pupd-err}
  {}
  { \pupd[\mask]{(\Exists \err. 0<\err \sep \upto{\err})} }
\end{mathpar}

The key novelty of the probabilistic modality is its ability to populate presampling tapes as shown in \ruleref{pupd-presample-exp}.
The rule says that if we own a presample tape we can populate the tape with a freshly sampled value $n$.
Similar to \ruleref{ht-rand-exp}, it allows re-distributing error credits along different branches of the randomized outcome, as long as the expected value of the error credit does not increase.
We showcase the probabilistic update modality on an example in \cref{sec:prove-implementation-counter}.

As a somewhat orthogonal property, the probabilistic update modality also internalizes the notion of continuity of probabilities within the logic.
Specifically, it permits synthesizing some arbitrarily small error credit \emph{out of thin air} as seen in \ruleref{pupd-err}.
This principle enables \emph{induction by error amplification} \cite{eris} as we showcase in \cref{sec:prove-implementation-counter}.

\section{Modular Specifications of Concurrent Randomized Modules}
\label{sec:modularity}

In this section, we first provide an overview of how HOCAP-style specifications capture logically atomicity of concurrent data structures.
Next, we explain how we extend the approach to capture \emph{randomized} logical atomicity and present a modular specification for the randomized counter module.
We also describe how the specification is strong enough to verify clients that use the randomized counter module concurrently.
Subsequently, we present three different implementations of the concurrent randomized counter module and discuss how to show that they satisfy the specification.
Later in \cref{sec:case-studies} we discuss how we verify a series of larger case studies.
\subsection{Modular Specifications of Concurrent Randomized Modules: Overview}
\label{sec:tech-overview-modularity}

Before considering the randomized setting, we showcase our specification style on a non-randomized example: a concurrent counter module with functions for creating a counter, (deterministically) incrementing by one, and reading.

As alluded to in \cref{subsec:rules}, the high level idea is to parameterize specifications by a view shift that captures how the logical state of the counter evolves at the linearization point.
Our specification of the non-randomized counter module is shown in \Cref{fig:basic-counter-spec}.
\begin{figure}[h!]
  \centering
  {   
    \begin{align*}
  & \hoare
    {\TRUE}
    {\createcounter\TT}
    {\Ret c . \Exists \iname . \counterpred~\iname~c \sep \cfrag~1~0}  \\
  & \All \mask, \iname, c, \propB .
    \hoare
    {\counterpred~\iname~c \sep (\All z. \cauth\ z \wand \pvs[\mask] \cauth\ (z+1) \sep \propB~z )}
    {\incrcounter~c}
    {z \ldotp Q\ z}[\mask \uplus \{\iname\}]  \\
  & \All \mask, \iname, c , \propB.
    \hoare
    {\counterpred~\iname~c \sep (\All z. \cauth\ z \wand \pvs[\mask] \cauth\ z \sep \propB~z )}
    {\readcounter~c}
    {z \ldotp \propB~z}[\mask \uplus \{\iname\}]
    \end{align*}
    }
  \caption{Specification for a (non-randomized) concurrent counter module.}
  \label{fig:basic-counter-spec}
\end{figure}

When creating a counter, one obtains ownership of two resources: $\counterpred~\iname~c$ and $\cfrag~1~0$.
The $\counterpred~\iname~c$ resource captures that $c$ is a counter with an associated invariant name $\iname$.
Intuitively, this invariant  contains the internal state of the counter but the details are unknown to clients.
The predicate is persistent, \ie{}, $\counterpred~\iname~c \provesIff \counterpred~\iname~c\sep \counterpred~\iname~c$ and can hence be freely shared.

The predicates $\cauth$ and $\cfrag$ provide \emph{authoritative} and \emph{fragmental} views of the counter.
Intuitively, $\cauth$ provides the counter module's view of the counter and $\cfrag$ denotes the client's view.
A fragmental view $\cfrag~q~n$ denotes a $q$-fractional view that the counter is \emph{at least} the value $n$.
The $\cfrag~q~n$ resource can be split and combined, \ie{}, $\cfrag~(q_{1} + q_{2})~(n_{1} + n_{2}) \provesIff \cfrag~(q_{1},n_{1}) \sep \cfrag(q_2, n_{2})$ and thus shared.
The fragmental view is guaranteed to be consistent with the authoritative view, \ie{}, $\cauth~n \sep \cfrag~q~m \proves m \leq n$ and $\cauth~n \sep \cfrag~1~m \proves m = n$, and updated accordingly, \ie{}, $\cauth~n\sep\cfrag~q~m \vdash \pvs[\mask] \cauth~(n + p)\sep\cfrag~q~(m + p)$.

The specification for the increment and read functions are parameterized by a view shift that gives (temporary) access to the module's view.
This is one of the key ideas of HOCAP-style specifications.
From the client's perspective, the view shift is a proof obligation.
For the increment function, proving this view shift requires having ownership of a fragmental view (to update the resources), but the fragmental view can be provided by opening an invariant using the update modality.
The client-chosen predicate $\propB$ lets the client derive information as part of the view shift.
For example, they can pick $Q~z \eqdef{} \cfrag~q~(n + 1) \land z = (n +1)$.

\paragraph{Probabilistic Concurrent modules with Error Redistribution.}
Now, consider the randomized concurrent counter module from \cref{sec:tech-overview} where the increment function increments the counter by a value chosen uniformly at random from 0 to 3.
For the client to be able to redistribute error credits as part of the random sampling, we parameterize the specification of the increment function by another view shift as shown in \cref{fig:rand-counter-spec-with-errors}.

\begin{figure}[h!]
  \centering
{  
  \begin{align*}
   \All \mask, \iname, c, \propB.
    \spac \hoareV {
    \counterpred~\iname~c\sep
    \pvs[\mask][\emptyset]
    \left(
    \begin{array}{l}
      \Exists \err, \Err .
      \begin{aligned}[t]
        & \upto{\err} \sep (\expect[\unifd{3}]{\Err} \leq \err) \sep \All x \in \intrange{0}{3}. \upto{\Err(x)} \wand{} \\
        & \big(\pvs[\emptyset][\mask](\All z. \cauth\ z \wand \pvs[\mask] \cauth\ (z+x) \sep Q\ \err\ \Err\ x\ z) \big)
      \end{aligned}
    \end{array}
    \right)
    }{\incrcounter\ c}{ \Ret z . \Exists \err, \Err, x . Q\ \err\ \Err\ x\ z}[\mask \uplus \{\iname\}]
\end{align*}}
    \caption{Specification of $\incrcounter$ for a concurrent randomized counter module.}
    \label{fig:rand-counter-spec-with-errors}
\end{figure}
Notice that in the precondition the client now has to prove a view shift which is split into two parts.
We begin by looking at the second part (the line at the bottom).
This is analogous to the deterministic case, except that the abstract state $\cauth~z$ gets incremented by some uniformly sampled $x \in \intrange{0}{3}$.
This operation is randomized, so we also let the client update their error credits along this distribution, which is the first part of the view shift.
After opening all invariants in $\mask$, the client chooses some $\err$ and an error distribution function $\Err$, gives up $\upto{\err}$, gets back $\upto{\Err(x)}$, re-establishes all invariants in $\mask$, and goes on to prove the second part.
Notice that the specification allows the client to retrieve error credits from an invariant.
Intuitively, these two parts of the view shift capture two separately logically atomic actions of the increment operation.
The first being the random operation where we re-distribute errors, and the second being the actual increment, where we increase the counter by the sampled value.
If all these preconditions are satisfied, then at the end of $\incrcounter$, we return some value $z$ which satisfies $\propB\ \err\ \Err\ x\ z$ for some $\err$, $\Err$, and $x$.
The specification for creating and reading the counter are unchanged as no randomization is involved.

\paragraph{Probabilistic Concurrent Modules with Error Redistribution and Presampling.}

One limitation of the previous specification is that the sampling operation is fixed to take place within the function call $\incrcounter$.
As a result, the only point at which randomness can be generated for the module, and errors can be distributed, is at the invocation of the increment operation.
However, it is sometimes useful to reason about the probabilistic part of the operation asynchronously.

In Clutch~\cite{clutch} presampling tapes are used to generate randomness
asynchronously and facilitate refinement proofs. In a concurrent setting, there
is also an asynchronous component arising from the order in which randomized
operations are physically resolved, and we propose the use of presampling and
tapes to resolve them in advance and independently from this order.

In the previous specification, the view shift consisted of a probabilistic part (\ie{}, spending and distribution of error credits) and a deterministic part (updating the abstract state).
Presampling allows us to decouple these two parts and reason about them separately, resulting in a more expressive HOCAP-style specification.
We demonstrate that by exposing tapes
and presampling operations in the module specifications, clients can perform
presampling for an abstract randomized operation. This is an \emph{indispensable} technique
for verifying certain concurrent modules, and we show an example in \cref{sec:spec-errors-tapes-proof}.

The new and final specification of the probabilistic
counter module, which includes not only error redistribution, but also
a (ghost) method for creating an abstract presampling tape and a (ghost) operation for sampling
on a tape, see \Cref{fig:rand-counter-spec-with-tapes}.%
\begin{figure}[h!]
  \centering
  \small
\begin{align*}
  & {\All \mask, \err, \Err, \tape, \lbl .
        \mhl{(\upto{\err} \sep (\expect[\unifd{3}]{\Err} \leq \err) \sep \ctape\ \lbl\ \tape \wand
        \spac \pupd[\mask]{\Exists n\in\intrange{0}{3}. \upto{\Err(n)} \sep \ctape\ \lbl\ (\tape\lapp [n])} )}
    } \\[1mm]
  & {\All \iname, c.
    \hoare
    {\counterpred~\iname~c}
    {\createtape\TT}
    {\Ret \lbl . \ctape\ \lbl\ \nil} } \\[1mm]
  & \All \mask, \iname, c, n, \tape, \propB.
    \spac \hoare {
    \begin{array}{l}
      \counterpred~\iname~c\sep { \mhl{\ctape\ \lbl\ (n\cons\tape)}} \sep \\
      \spac(\All z. \cauth\ z \wand \pvs[\mask] \cauth\ (z+n) \sep Q\ z)
    \end{array}
    }{\incrcounter\ c\ \lbl}{z. \mhl{\ctape\ \lbl\ \tape} \sep Q\ z}[\mask \uplus \{\iname\}]
\end{align*}
\caption{Specification for a randomized counter module with presampling tapes.}
\label{fig:rand-counter-spec-with-tapes}
\end{figure}
This specification has a new predicate $\ctape$ that stores the presampled
randomness for the random counter. Note that $\ctape$ is an abstract predicate
which might be realized in multiple ways besides using primitive tape predicates,
which allows us to hide the details of how different implementations of the counter
module physically generate randomness.
By exposing the abstract presampling tape explicitly, we aim to capture more of the proof principles for a concrete randomized operation.
(In \cref{sec:spec-errors-tapes-proof}, we demonstrate that this specification which exposes
abstract tapes is in fact \emph{more general} than the previous one)

Compared to the previous randomized specification, reasoning about randomness of the increment operation is now extracted into a separate condition that utilizes the probabilistic update modality ($\pupd\prop$), which says that we can presample onto the $\ctape$ and distribute errors in a expectation-preserving manner. With this change, clients can allocate their own local tapes via $\createtape$ and reason about randomness locally. %
The $\incrcounter$ function takes a non-empty $\ctape$ predicate as argument, and acts in a (logically) deterministic manner, by reading and consuming the first element $n$ of the tape, and incrementing the abstract state of the counter by $n$.

Now that we have shown an expressive specification (\cref{fig:rand-counter-spec-with-tapes}), in the following sections, we show how this specification suffices to verify clients.
We also show that three different implementations of the probabilistic random counter module all meet this specification.
These three implementations exhibit different numbers of sampling operations, but yet they all meet the same abstract module specification.
In other words, the randomization of the increment operation acts ``logically atomic'' as expressed by a single probabilistic update, even if in reality, it is not.
From the perspective of a client, random sampling within the increment operation appears to behave as if it is simply a single $\Rand 3$.
We refer to this as \emph{randomized logical atomicity}.

\subsection{Verifying Clients of Randomized Counter Module}
\label{sec:spec-errors-tapes-proof}
We now describe how the specification with error re-distribution and presampling tapes shown in \Cref{fig:rand-counter-spec-with-tapes} can be used to verify concurrent clients. We also show how the HOCAP-style specification that exposes abstract tapes is more general than the one that does not.

\paragraph{Revisiting $\contwodie$}
We begin with the $\contwodie$ client example introduced in \cref{sec:tech-overview}.
Since the new specification of the randomized concurrent counter utilizes tapes, we annotate the $\contwodie$ client to use the abstract tapes exposed in the specification:
\begin{align*}
  \contwodie \eqdef{}
  & \Let c = \createcounter\TT in  \\
  &\left({
    \begin{array}{l}
      \Let \lbl=\createtape \TT in\\
      \incrcounter\ c\ \lbl\\
    \end{array}
    }
    \middle|\middle|\middle|
    {\begin{array}{l}
      \Let \lbl=\createtape \TT in\\
      \incrcounter\ c\ \lbl\\
    \end{array}}\right); \\
  &\readcounter\ c
\end{align*}
Recall that we expect the return value to be $0$ with a probability $1/16$.
We state this through the following \theaplog Hoare triple: $\hoare{\upto{1/16}}{\contwodie}{v.v>0}$.

We present here a high level intuition for the proof and defer the details to \appref{sec:app-proof-counter-client}.
Most of this proof is similar to the one sketched in \cref{sec:tech-overview} where we allocate an invariant that encodes a protocol that tracks both the available amount of error credits and the ghost state of both threads and describes how they can evolve.
In the case where both threads sampled $0$, we are able to obtain $\upto{1}$ from the invariant at the end and derive a contradiction with \ruleref{err-1}.

The difference between this proof and that from \cref{sec:tech-overview} is twofold.
Firstly, the randomness is generated asynchronously using the presampling rule and the abstract tapes.
The probabilistic update modality allows us to open the invariant, obtain error credits from it, presample onto our abstract tapes, redistribute the error credits, and close the invariant again, all in an atomic manner.
Secondly, to apply $\incrcounter$ and $\readcounter$, we need to prove the view shifts in the precondition of their corresponding Hoare triple specifications.

An important detail of this proof is that we do not need to place any $\ctape$ predicate in the invariant.
Each thread uses a separate and local tape which does not need to be shared.
This kind of ``local tape'' principle lets each thread ``own'' its own randomness and this simplifies the proof since we need not worry how the state of $\ctape$ is changed by other external concurrent threads.

\paragraph{Advantage of Exposing Abstract Tapes}
Recall that in \cref{sec:tech-overview-modularity}, we presented a simpler specification of the randomized counter module (\cref{fig:rand-counter-spec-with-errors}) that does not expose presampling tapes as abstract predicates. 
To see why that specification is not as general as that in \cref{fig:rand-counter-spec-with-tapes} and that it is useful to expose the presampling tapes in the specification, consider the following $\twoincr$ program and its specification in \cref{fig:twoincr}.
\begin{figure}[htbp]
  \centering
   \begin{subfigure}[t]{0.45\textwidth}
              \begin{align*}
                \twoincr\ \_\eqdef{}& \Let c = \createcounter\ \TT in \\
                & \Let \lbl = \createtape\ \TT in \\
                & \incrcounter\ c\ \lbl; \\
                & \Let v_1 = \readcounter\ c in\\
                & \incrcounter\ c\ \lbl; \\
                & \Let v_2 = \readcounter\ c-v_1 in\\
                & 4 \cdot v_1 + v_2
              \end{align*} 
   \end{subfigure}
            ~
   \begin{subfigure}[t]{0.45\textwidth}
     \begin{align*}
       \hoareV {
         \begin{array}{l}
           \pvs[\mask][\emptyset] \Exists \err, \Err. 
           \upto{\err} \sep (\expect[\unifd{15}]{\Err} \leq \err) \sep \\
           \qquad\; (\All x. \upto{\Err(x)} \wand \pvs[\emptyset][\mask] \propB\ \err\ \Err\ x)
      \end{array}
       }{\twoincr\ \TT}{z \ldotp \Exists \err, \Err . Q\ \err\ \Err\ z}[\mask\uplus\{\iname\}]
     \end{align*}
  \end{subfigure}
            \caption{Implementation and specification of $\twoincr$.}
            \label{fig:twoincr}
\end{figure}

The sequential program $\twoincr$ first creates a new randomized counter and allocates a tape for the counter.
It then performs two $\incrcounter$ and $\readcounter$ pair operations successively, to read the exact values $v_1$ and $v_2$ added to the counter.
At the end it returns $4 \cdot v_1+v_2$.
As both $v_1$ and $v_2$ are sampled uniformly from $\{0,\dots,3\}$, the return value is uniformly distributed between $\{0,\dots,15\}$.
This is captured by the Hoare triple in \cref{fig:twoincr} where error credits can be re-distributed across the $16$ possibilities in an expectation-preserving way.
Note that the view shift of the Hoare triple captures the fact that the re-distribution happens in a logically-atomic manner.

Proving the specification of $\twoincr$ with the more general specification (\cref{fig:rand-counter-spec-with-tapes}) is relatively straightforward. After applying the specification for creating the counter and the tape, we perform two consecutive presamples onto $\ctape$ with the presampling specification of the counter module. These two presampling operations are combined into one atomic operation with the \ruleref{pupd-bind} rule, allowing us to use the view shift provided in the precondition to split the error credits for the $16$ possibilities. The rest of the proof follows directly by applying the specification for incrementing the counter with the tape and reading from it twice.

However, the specification without the presampling tapes exposed (\cref{fig:rand-counter-spec-with-errors}) is not strong enough to prove this Hoare triple.
The specification restricts the error redistribution to only occur within the $\incrcounter$ call, and we are unable to combine the two separate error redistribution operations in each $\incrcounter$ call into one atomic action.
On the other hand, the more general specification allows us to ``pull'' the randomized operation out of the $\incrcounter$ call and perform the randomized operation in advance using the presampling operation of the abstract tapes.

\subsection{Three Implementations of the Randomized Counter Module}
\label{sec:three-implementations}
Recall from \Cref{sec:tech-overview-modularity} that the specification of the randomized counter module from \cref{fig:rand-counter-spec-with-tapes} provides four methods: $\createcounter$ for creating the counter, $\createtape$ for creating a tape, $\incrcounter$ for incrementing the counter with a random value chosen uniformly from the set $\{0, \ldots, 3 \}$ (sampled from the tape), and $\readcounter$ for reading the value of the counter.

To illustrate the expressiveness of our modular specification, we consider three implementations that we show meet the same specification, which we refer to as $I_{1}$, $I_{2}$, and $I_{3}$, respectively.
They only differ in the way they implement the $\createtape$ and $\incrcounter$ method--the implementations of $\createcounter$ and $\readcounter$ are the same in all three implementations:
\[
  \createcounter \eqdef{} \Lam \_ . \Alloc 0  \qquad
  \readcounter \eqdef{} \Lam l . \deref l
\]
Internally, the counter is represented by a pointer to a number and the read method simply dereferences the pointer.
The three implementations of the create tape and increment methods are shown in \Cref{fig:implementation-counter}.
\begin{figure}[htbp]
  \centering
   \begin{subfigure}[t]{0.4\textwidth}
              \begin{align*}
                \createtape_1 \eqdef{} & \Lam \TT. \AllocTape 3\\
                \createtape_2 \eqdef{} & \Lam \TT. \AllocTape 1\\
                \createtape_3 \eqdef{} & \Lam \TT. \AllocTape 4
              \end{align*}
   \end{subfigure}
            ~
  \begin{subfigure}[t]{0.5\textwidth}
              \begin{align*}
                \incrcounter_1 \eqdef{} & \Lam l, \lbl. \Faa l\ (\Rand \lbl~3) \\
                \incrcounter_2 \eqdef{} & \Lam l, \lbl.
                                          \begin{aligned}[t]
                                            &\Let \lbl = \AllocTape 1 in \\
                                            & \spac \Faa l\ (\Rand \lbl\ 1 \cdot 2 + \Rand \lbl\ 1)
                                          \end{aligned} \\
                \incrcounter_3 \eqdef{} & \Rec f l\ \lbl= \DumbLet x = \Rand\lbl\ 4 \In \\
                & \phantom{\Rec f l\ \lbl = }~\If x<4 then \Faa l\ x \Else f\ l\ \lbl
              \end{align*}
  \end{subfigure}
            \caption{Implementation of the counter module.}
            \label{fig:implementation-counter}
\end{figure}
In $I_{1}$, the increment method simply increments the counter value stored at the location by a $\Rand 3$ chosen value between $0$ and $3$ using a fetch-and-add instruction.
The function hence creates a tape with bound $3$.
In $I_{2}$, the increment method is implemented using two coin flips (\ie{}, calls to $\Rand 1$), and in $I_{3}$, we use a recursive rejection sampler that, in order to simulate $\Rand 3$, repeatedly samples from $\Rand 4$ until it gets a value within $\{0, \ldots 3 \}$.
The $\createtape$ function for both implementations creates a tape with bound $1$ and $4$, respectively.

For $I_{2}$ and $I_{3}$ in particular, it is interesting that even though the implementations do not sample randomness atomically (e.g.,~$I_3$ can possibly execute any number of $\Rand 4$ operations), they still meet the specification where the presampling of a single value onto the abstract tape is described by a \emph{single} probabilistic update modality as we show in the next section.
In other words, we capture \emph{randomized logically atomicity} of the module in the sense that externally, there appears to be a single randomized transition within the $\incrcounter$ function.

\subsection{Verifying $I_{1}$, $I_{2}$, and $I_{3}$}
\label{sec:prove-implementation-counter}
We now show how the three randomized counter implementations meet the specification with error redistribution and presampling tapes.
We start by giving concrete definitions for the three abstract predicates.
For the three implementations, it turns out that the counter predicate $\counterpred$, and the $\cauth$ and $\cfrag$ predicates are defined identically; the persistent counter predicate $\counterpred~\iname~c$ is defined as $\knowInv{\iname}{\Exists l, n.
  c=l \sep l\mapsto n \sep \cauth\ n}$ and the $\cauth$ and $\cfrag$ predicates are defined with a standard authoritative-fractional resource algebra \cite{irisjournal}.
We show the exact definition of $\ctape$ for each of the three implementations below.
\newcommand{\expand}{\textlog{expand}}
\begin{align*}
  \ctape_1\ \lbl\ \tape \eqdef{} & \progtape{\lbl}{3}{\tape}\\
  \ctape_2\ \lbl\ \tape \eqdef{} & \progtape{\lbl}{1}{\expand\ \tape}\sep (\All x \in \tape . x < 4)  \\
  \ctape_3\ \lbl\ \tape \eqdef{} & \Exists \tapeB. \filter\ (\Lam x.x<4)\ \tapeB = \tape \sep \progtape{\lbl}{4}{\tapeB}
\end{align*}
For $\ctape_1$, since the first implementation uses a $\Rand 3$ to sample from $0$ to $3$ directly, we define $\ctape_1$ with the presampling tape $\progtape{\lbl}{3}{\tape}$.
For $\ctape_2$, since we are sampling from $0$ to $3$ via two $\Rand 1$s, the predicate is defined by expanding the tape elements into its binary representation.
The function $\expand$ takes in a list of numbers and rewrites them into binary representation while keeping the list ``flattened''.
For example $\expand ([2;3;1;0])$ returns $[1;0;1;1;0;1;0;0]$.
Finally, for the third implementation, if we are logically storing $\tape$ with our $\ctape$ predicate, the concrete tape stores some list $\tapeB$ such that $\tape$ is equal to $\tapeB$ with all $4$s removed from it.

It suffices to show that the functions $\createcounter$, $\createtape$, $\incrcounter$, and $\readcounter$ satisfy the specification and that $\ctape$ satisfies the presampling probabilistic update specification, \ie{}, we can logically append a new element into the $\ctape$ while redistributing errors.
The specification of the functions are not too complicated.
As an example, consider the $\incrcounter$ specification for $I_3$.
\[
  \hoareV {
    \counterpred~\iname~c\sep {\ctape\ \lbl\ (n\cons\tape)} \sep
    (\All z. \cauth\ z \wand \pvs[\mask] \cauth\ (z+n) \sep Q\ z)
}{\begin{array}{l}
    \incrcounter_3~c~\lbl
  \end{array}
}{\Ret z . {\ctape\ \lbl\ \tape} \sep Q\ z}[\mask \uplus \{\iname\}]
\]
After unfolding the definition of the abstract predicates for $I_3$, we repeatedly loop through the recursive function until we reach a value $n$ in the tape that is smaller than $4$ by structural induction on the tape or Löb induction.
During the atomic $\langkw{faa}$ operation, we open the invariant with \ruleref{ht-inv-open} and eliminate the view shift in the precondition.
The specification of the other functions can be proven similarly.

We now focus on showing that for each of the $\ctape$ definitions, they satisfy the presampling specification.
For $\ctape_1$, we see after unfolding its definition, the statement of the presampling specification is the same as that of \ruleref{pupd-presample-exp} and hence holds directly.
For $\ctape_2$, it suffices to prove the following probabilistic update:
\begin{equation}
  \label{eq:presample-spec2}
  \begin{split}
    &(\expect[\unifd{3}]{\Err} \leq \err) \wand
      \progtape{\lbl}{1}{\tape} \wand
  \upto{\err} \wand \\
  &\spac \pupd[\mask]{\Exists v_1, v_2. \upto{\Err(v_1 \cdot 2+v_2)} \sep \progtape{\lbl}{1}{\tape \lapp [v_1, v_2]}}
  \end{split}
\end{equation}
This probabilistic update is valid because we can do two presamples consecutively via \ruleref{pupd-bind}.
We first apply \ruleref{pupd-presample-exp} to presample the first bit, choosing the first error splitting function $\Err_a \eqdef{} \Lam b.
\text{if } b=1 \text{ then } \Err(2)+\Err(3) \text{ else } \Err(0)+\Err(1)$.
We then do a case distinction on the bit that was sampled.
If it is $0$, we apply \ruleref{pupd-presample-exp} again, choosing the error splitting function to be $\Err_b\eqdef{} \Lam b. \text{if } b=1 \text{ then } \Err(1) \text{ else } \Err(0)$.
Otherwise, we choose $\Err_b\eqdef{} \Lam b. \text{if } b=1 \text{ then } \Err(3) \text{ else } \Err(2)$.

For $\ctape_3$ we want to show that we can repeatedly presample enough values onto the tape such that the last element is smaller than $4$ and all values beforehand are $4$, while distributing the error credit according to the final value.
This can be shown by the following lemma:
\begin{equation}
  \begin{split}
   &(\expect[\unifd{3}]{\Err} \leq \err) \wand (\Exists \tapeB.\filter\ (\Lam x. x<4)\ \tapeB =\tape\sep\progtape{\lbl}{4}{\tapeB}) \wand
  \upto{\err} \wand \\
  &\spac \pupd[\mask]{\Exists n. 0\leq n<4 \sep \upto{\Err(n)} \sep \Exists \tapeB.\filter\ (\Lam x. x<4)\ \tapeB =(\tape\lapp[n])\sep\progtape{\lbl}{4}{\tapeB}}
  \end{split}
  \label{eq:presample-spec3}
\end{equation}
We prove this probabilistic update through \emph{induction by error amplification}.
We first apply \ruleref{pupd-bind} and \ruleref{pupd-err} to obtain some positive error credit $\upto{\err'}$ to get the following:
\begin{align*}
  &(\expect[\unifd{3}]{\Err} \leq \err) \wand
  \spac \mhl{\err'>0 \wand \upto{\err'}\wand}
   \spac (\Exists \tapeB. \filter\ (\Lam x. x<4)\ \tapeB =\tape\sep \progtape{\lbl}{4}{\tapeB}) \wand
  \upto{\err} \wand \\
  &\spac \pupd[\mask]{\Exists n. 0\leq n<4 \sep \upto{\Err(n)} \sep (\Exists \tapeB.\filter\ (\Lam x. x<4)\ \tapeB =(\tape\lapp[n])\sep\progtape{\lbl}{4}{\tapeB})}
\end{align*}

Now we apply the induction by error amplification rule below (see Eris~\cite{eris} for more details):
\[
     \inferH{ind-err-amp}
     {\err_1 > 0 \\  k > 1 \\ \upto{\err_1} \\ \All \err_2. (\upto{k\cdot \err_2} \wand \prop) \wand \upto{\err_2} \wand \prop }
     { \prop}
\]
Morally this states that in order to prove $P$ we can assume it holds guarded by an amount of credits amplified
by a factor $k$ strictly greater than 1.
We choose the amplification factor $k\eqdef{}5$.
It suffices to show (with the induction hypothesis highlighted):
\begin{align*}
  & (\expect[\unifd{3}]{\Err} \leq \err) \wand \err'>0\wand  \mhl{\big(\upto{5 \cdot \err'}\wand
     (\Exists \tapeB.\filter\ (\Lam x. x<4)\ \tapeB =\tape \sep\progtape{\lbl}{4}{\tapeB})\; \dots \big) \wand} \\
  & \spac \spac \upto{\err'}\wand
     (\Exists \tapeB.\filter\ (\Lam x. x<4)\ \tapeB =\tape\sep\progtape{\lbl}{4}{\tapeB}) \wand
  \upto{\err} \wand \dots
\end{align*}
We can now combine $\upto{\err}\sep\upto{\err'}$ with \ruleref{err-split} and apply \ruleref{pupd-presample-exp} with $\upto{\err+\err'}$ as the initial error budget.
We choose the distribution function to be $\Lam x. \text{if } x<4 \text{ then } \Err (x) \text{ else } \err+ 5 \cdot \err'$.
After a single presampling step, we do a case distinction on whether the presampled value is $4$ or not.
If it is, then we establish the conclusion with the induction hypothesis since we successfully amplified the error credit $\upto{\err'}$ by a factor of $5$.
Otherwise, we presampled an ``accepted'' value, and we can directly establish the goal via \ruleref{pupd-ret}.

\section{Case Studies}
\label{sec:case-studies}
In this section, we present several other case studies that we have verified using \theaplog.

\subsection{Thread-Safe Hash Functions}
\label{sec:concurrent-hash}
\newcommand{\conhashfun}{\textsf{hashFun}~\gamma}
\newcommand{\conhashkey}{\textsf{hashKey}~\gamma}
\newcommand{\initconhash}{\textlang{hashInit}}
Hash functions are often assumed to behave \emph{uniformly}~\cite{uniform-hash-assumption}.
That is, a hash function $h$ from a set of keys $K$ to a set of values $V$ behaves as if, for each key $k$, the hash $h(k)$ is randomly sampled from a uniform distribution over $V$ independently of all other keys.
This assumption can be modeled using an idealized hash function that uses a mutable map, which serves as a cache of hashes computed so far \cite{DBLP:series/isc/MittelbachF21}.
If the key has already been hashed, we return the value stored in the map, otherwise we sample a fresh value uniformly, store it in the cache, and return it.
In the concurrent setting, however, this does not suffice: if two threads concurrently attempt to hash the same key $k$, they may end up with different hash values.
If one thread gets preempted by the scheduler right after sampling, a second thread could overtake and sample a different value before the first thread stores its value to the cache.

We implement a thread-safe idealized hash function using a lock.
To hash a value, one first acquires the lock, then samples the key and stores it to the cache, before releasing the lock again.
While the implementation is uninteresting, its specification is not.
In particular, we give a specification that offers exclusive ownership of each key $k$ and the ability to presample the hash $h(k)$.
As we later see in \cref{sec:bloom-filter}, this ability can greatly simplify the probabilistic analysis of concurrent data structures that use hashing.

The $\initconhash$ function initializes a new hash function and satisfies the specification below.
\begin{align*}
  \hoare{\TRUE}
  {\initconhash~()}
  { h \ldotp \Exists \gamma. \conhashfun~h \sep \textstyle\Sep_{k \in K} \conhashkey~k~-}
\end{align*}
Here, $\gamma$ is a ghost name logically identifying the hash function.
The abstract predicate $\conhashfun~h$ is duplicable, \ie{}, $\conhashfun~h \provesIff \conhashfun~h \sep \conhashfun~h$, while $\conhashkey~k~-$ represents that key $k$ has not yet been hashed, and is exclusive, \ie{}, $\conhashkey~k~- \sep \conhashkey~k~- \proves \FALSE$.
When invoking a hash function on a key with an undecided value, a fresh value $v \in V$ is sampled and $\conhashkey~k~v$ is returned.
The predicate $\conhashkey~k~v$ is duplicable and each subsequent invocation is guaranteed to return $v$.
\begin{align*}
  &\hoare
  {\conhashfun~h \sep \conhashkey~k~- }
  {h~k}
  {\val \ldotp \Exists v \in V . \conhashkey~k~v} \\
  &\hoare
  {\conhashfun~h \sep \conhashkey~k~v }
  {h~k}
  {\valB \ldotp \valB = \val}
\end{align*}
However, hash values can also be presampled and error credits redistributed across the possible outcomes of the presampling using the probabilistic update below.
\begin{align*}
  &\conhashfun~f \sep \conhashkey~k~- \sep \upto{\err} \wand{} \\
  &\pupd[\top] \Exists v \in V . \conhashkey~k~v \sep (v \in X \sep \upto{\err_{1}}) \lor (v \not\in X \sep \upto{\err_{0}})
\end{align*}
Here $X \subseteq V$ is some set of hash values and $\err_{1}, \err_{0} \in [0,1] $ such that $\err_{1} \cdot |X| + \err_{0} \cdot (|V| - |X|) \leq \err \cdot |V|$.
For example, by picking $\err_{1} \eqdef{} 1$, $\err_{0} \eqdef 0$, and $\err \eqdef{} {|X|}/{|V|}$ one can spend $\err$ error credits to avoid the outcomes in $X$ when determining the hash $h(k)$.

We show the specification by allocating a fresh presampling tape for each key in K.
A similar idea is used in previous work \cite{clutch} to show refinement of lazy and eager hash functions.
In our specification, intuitively, $\conhashkey~k~-$ denotes exclusive ownership of $k$'s presampling tape which is transferred to an invariant after presampling.
This invariant captures that, for all keys $k$, \emph{either} $n$ has been presampled onto $k$'s tape \emph{or} $n$ has been stored at entry $k$ in the hash function's cache.

\subsection{Bloom Filter}
\label{sec:bloom-filter}

\begin{figure}[tbp]
  \small
   \begin{subfigure}[t]{0.43\textwidth}
              \begin{align*}
                      &\bfinit\ ()\eqdef{} \\
                      & \qquad \Let \langv{hfs} = \langv{List.init}~k~(\Fun \_. \langv{hash\_init}~()) in  \\
                      & \qquad \Let \langv{arr} = \langv{Array.init}~S~\False in \\
                      & \qquad (\langv{hfs},\langv{arr})\\
              \end{align*}
   \end{subfigure}
    \begin{subfigure}[t]{0.43\textwidth}
 \begin{align*}
        & \bfinsert\ \langv{bfl}\ x \eqdef{} \\
        & \qquad \Let (\langv{hfs},\langv{arr}) = \langv{bfl} in \\
        & \qquad \langv{List.Iter}(\Fun h. \Let i = h~x in \\
        & \qquad \hspace{5.45em} \langv{arr}[i] \gets \True)~\langv{hfs}
 \end{align*}
 \end{subfigure}
   \begin{subfigure}[t]{0.5\textwidth}
\begin{align*}
        & \bflookup\ \langv{bfl}\ y \eqdef{} \\
        & \qquad \Let (\langv{hfs},\langv{arr}) = \langv{bfl} in \\
        & \qquad \Let \langv{res} = \Alloc \True in \\
        & \qquad \langv{List.Iter}(\Fun h. \Let i = h~y in \\
        & \hspace{7.6em} \langv{res} \gets \deref{\langv{res}} \mathop{\&\&} \langv{arr}[i])~\langv{hfs}; \\
        & \qquad \deref{\langv{res}}
\end{align*}
   \end{subfigure}
   \begin{subfigure}[t]{0.4\textwidth}
\begin{align*}
        & \bfmain\ \langv{xs}\  \langv{y}\eqdef{} \\
        & \qquad \Let \langv{bfl} = \bfinit\ () in \\
       & \qquad (\Rec \langv{f}~\langv{zs}~ = \\
        &\qquad \hspace{1em} \langkw{match}\ \langv{zs}\ \langkw{with} \\
        & \qquad \hspace{1em} \spac|\ [~] \Ra () \\
        & \qquad \hspace{1em} \spac|\ \langv{z} :: \langv{zs'} \Ra \Parallel{(\bfinsert\ \langv{bfl}\ \langv{z})}{(\langv{f}~\langv{zs'})} \\
        & \qquad \hspace{1em}  \langkw{end})~\langv{ks};\\
        & \qquad \bflookup\ \langv{bfl}\ \langv{k}
\end{align*}
   \end{subfigure}
  \caption{Implementation of a concurrent Bloom filter. }
  \label{fig:bfilter-code}
\end{figure}

Bloom filters are approximate data structures to represent sets, with operations for inserting elements and querying for membership.
In their most basic, sequential presentation, a Bloom filter consists of an array of bits of a fixed size $S$, initially set to 0, and a list of hash functions $(h_1, \dots, h_k)$ of some fixed length $k$.
When inserting an element $x$, we compute $(h_1(x)\!\mod S, \dots, h_k(x)\!\mod S)$ and set those indices to 1.
When checking if an element $y$ is in the set, we also compute $(h_1(y)\!\mod S, \dots, h_k(y)\!\mod S)$, and look up those indices in the array.
If they are all set to 1, we answer positively, otherwise we answer negatively.
Thus, when checking the membership of an element that is not in the set there exists a small probability of observing a false positive if there are hash collisions with previously inserted elements.
Computing this probability is challenging and requires involved combinatorial reasoning, in fact Bloom's original analysis~\cite{Bloom70} gave the wrong bound.

An efficient concurrent implementation of a Bloom filter allows parallel
insertions, since concurrent writes to the same entry in the array  would both set the entry to 1.
In this case study, we implement
a concurrent Bloom filter and prove a bound on the probability of observing a
false positive result on a membership query. We use the concurrent hash module
presented in \cref{sec:concurrent-hash} to implement this concurrent Bloom filter
example (see \cref{fig:bfilter-code}).

First consider $N$ sequential insertions $x_1,\dots,x_N$ followed by
checking membership for some $y \not\in \{x_1, \dots, x_N \}$.
From a mathematical perspective, the probability of false positive corresponds
to the following experiment: first sample a batch of $k\cdot N$ integers uniformly
at random in $\{0, \ldots, S-1\}$. Now sample a second batch of $N$ integers in
the same manner. What is the probability that they are \emph{all} in the first
batch? The exact bound was first calculated by~\citet{BoseGKMMMST08}, and in later
work, \citet{GopinathanS20} mechanized the proof.

Now suppose that the insertions $x_1,\dots,x_N$ happen in $N$ parallel threads.
Intuitively, concurrent implementations of Bloom filters should have the same
probability of false positive, since parallel queries to hash functions are
independent. Using our logic, we can make this intuition concrete, and prove
that the bound in the concurrent setting indeed corresponds to the sequential one.

Our modular approach allows us to simplify the mathematical reasoning within
the proof of the specification and defer all complex combinatiorial reasoning
to the meta-level. The proof crucially relies on both the stateful
representation of error probabilities (\ie{}, error credits) as well as the
notion of randomized logical atomicity, which allows us to presample all randomness in advance.

\newcommand{\fperror}[2]{{\sf E}_{\sf fp}(#1,#2)}
The key observation is that the probability of false positive follows a simple recurrence. Let
$\fperror{l}{b}$ be the probability of observing a false positive for a single membership query after setting $l$
uniformly selected indices to $1$ in an array that already contains $b$ bits set to $1$.
\begin{align*}
        \fperror{0}{b} &\eqdef{} (\tfrac{b}{S})^k \\
        \fperror{l+1}{b} &\eqdef{} \tfrac{b}{S} \cdot \fperror{l}{b} + \tfrac{S-b}{S} \cdot \fperror{l}{b+1}
\end{align*}
Our analysis can therefore assume that every time we hash, we start with $\upto{\fperror{l+1}{b}}$ for some $l$,
where $b$ is the number of distinct hash outputs that have been observed so far,
and then obtain either $\upto{\fperror{l}{b}}$ or $\upto{\fperror{l}{b+1}}$ depending on whether or not
the output of the hash is a new one or not. This means that the decision on
how to distribute credits can be done locally everytime we hash a new element.

With this in mind,
we can prove the following spec:
\begin{equation}
    \label{eq:bf-main}
  \hoare{\sf{NoDup}(xs) \ast y \not\in xs \ast \upto{\fperror{k\cdot|xs|}{0}}}
  {\bfmain\ xs\ y}
  {v \ldotp v = \False}
\end{equation}
\ie{}, the probability of false positive is at most $\fperror{k\cdot|xs|}{0}$, which corresponds
to the theoretical bound given by~\citet{BoseGKMMMST08} for the sequential setting\footnote{Note that their bound is
given as a closed mathematical expression and we have not mechanized that it corresponds to our recursive definition.}.
In order to simplify reasoning about concurrent hashing,
we presample the hash outcomes for every key in $xs$ in advance, using the hash specification in
\cref{sec:concurrent-hash}. It is at this point that most reasoning about probabilities
takes place, and that we do the distribution of error credits. After this phase, we 
have $\upto{\fperror{0}{B}}$ for some $B$, as well as predicates of the form $\conhashkey~k_i~v_{i}$
for every key and every hash, and we know that the set of presampled hash outcomes has cardinality
$B$. Then we execute all insertions, with an invariant that ensures that the array
never has more than $B$ elements set to 1. Finally, we can do a lookup, and use our error credits
$\upto{\fperror{0}{B}}$ to ensure that at least one of the indices we look up is set to 0, which
guarantees that the query returns $\False$.

To the best of our knowledge, we are the first to
prove a tight bound on the probability of false positives for a concurrent Bloom filter~\eqref{eq:bf-main}.
For more details on the analysis, we refer the reader to our \rocq development.

\subsection{Lazy Random Sampler}
\label{sec:lazy-rand}
In this section, we consider the implementation of a concurrent lazy one-shot random sampler.
This sampler is lazy in the sense that we only perform the sampling the first time the thunk is invoked and we store the result in a reference that is read from whenever the thunk is invoked again.

An excerpt of the implementation of the lazy random sampler is shown in \cref{fig:lrand-code}.
The function $\lazyrandinit$ creates a tuple containing a lock and a reference that points to $\None$.
When we call $\lazyrandf$ with the tuple and a tape label, we acquire the lock and load the value of what the location is pointing to.
If it is $\Some x$, we directly return $x$.
Otherwise, we sample $x$ from the tape with the function $\randf~\lbl$ from some Rand module and store it into the location.
Here the Rand module is some abstract module that samples $\{0,\dots,\tapebound\}$ uniformly from some abstract tape $\lbl$, where $\tapebound$ is some parameter fixed in advance, i.e.~$\randf~\lbl$ acts like a normal $\Rand N~\lbl$ (we provide more details in \appref{sec:rand-module}).
We additionally take in an extra argument $tid$ in $\lazyrandf$ and store it into the location together with the sampled value to also track the first thread id that succeeds in acquiring the lock and performing the actual randomized operation.
We release the lock right before we return from the function.
\begin{figure}[htbp]
   \begin{subfigure}[t]{0.43\textwidth}
     \begin{align*}
       \lazyrandinit \eqdef{} & \Fun \_.   \\
                              &\quad\Let \loc = \Alloc \None in  \\
                              &\quad \Let lo = \newlock\ \TT in \\
                              &\quad (lo,\loc)\\
     \end{align*} 
   \end{subfigure}
   ~
   \begin{subfigure}[t]{0.43\textwidth}
\begin{align*} 
  \lazyrandf\eqdef{} &\Fun (lo,\loc), \lbl, tid. \\
  &\quad\acquire\ lo; \\
  &\quad\DumbLet v = \langkw{match} \deref \loc\ \langkw{with}  \\
  &\qquad \spac |\ \Some x \Ra x \\
  &\qquad \spac |\ \None \Ra \\
  &\quad\qquad\spac\spac\Let x = (\randf\ \lbl, tid) in \\
  &\quad\qquad \spac\spac\loc \gets \Some x;  x\\
  &\quad  \langkw{end} \In \\
  &\quad \release\ lo; v
\end{align*}
   \end{subfigure}
  \caption{Implementation of a lazy random sampler.}
  \label{fig:lrand-code}
\end{figure}

To motivate the specification of this lazy random sampler module, consider a client program $\lazyrace$ that uses the module.
In this example, we set the parameter of the internal Rand module to be $1$, so $\randf$ samples uniformly between $0$ and $1$.
(The function $\lazyalloc$ in this example creates a tape for this lazy random sampler, and we omit the code for brevity.)
\begin{align*} 
  \lazyrace \eqdef{} &\Let r = \lazyrandinit\ \TT in  \\
  & \Parallel{(\lazyrandf\ r\ (\lazyalloc\ \TT)\ 0)}{(\lazyrandf\ r\ (\lazyalloc\ \TT)\ 1)}
\end{align*}
In the $\lazyrace$ program, we create a lazy random sampler and fork two
threads. Each thread attempts to sample from it but they pass a different $tid$ as the thread id argument. It should be the case that both threads return the same tuple value
$x=(x_1, x_2)$; intuitively, regardless of how the threads are scheduled, the 
thread that is executed last must read the value stored by the the thread that is executed first.
Consider the following
specification of $\lazyrace$ where for both return values of the threads, we
have $x_1=x_2$ with error probability $1/2$.
\[\hoare{\upto{1/2}}{\lazyrace}{v.\Exists n. v= ((n,n),(n,n))}\]
This is true morally because whichever threads gets scheduled first to perform
the sampling, we can use $\upto{1/2}$ to avoid sampling a value from $\randf$
that is different from the $tid$ passed, ensuring that the sampled value is
identical to the $tid$. However, there is some subtlety in the proof of this
Hoare triple. In particular, we cannot perform any presampling in advance of the
actual $\lazyrandf$. If we directly attempt to presample a value to each tape on
both threads (avoiding the corresponding $tid$), we need to pay up to
$\upto{3/4}$ error credits because we are doing two presampling calls, where
ideally, we should only need to do one. One might try to rewrite $\lazyrace$
such that both threads share the same tape, but this does not solve the
presampling problem directly. In particular, before either threads call
$\lazyrandf$, we do not know what value to sample onto the tape. Whatever value
is presampled, the scheduler can deliberately choose to schedule
the threads in a way such that the $tid$ of the winning thread does not match
the presampled value. In other words, we want to delay the operation of
presampling and perform it not before the $\lazyrandf$ call, but during it.

Given this observation, the specification of the lazy random sampler module is
written in a way that allows presampling to be performed within the $\lazyrandf$
call dynamically.
We show the specification for presampling and $\lazyrandf$ in \cref{fig:lazy-rand-spec}.
\begin{figure}[ht]
  \begin{subfigure}[t]{1\textwidth}
    \begin{align*}
      &(\expect[\unifd{\tapebound}]{\Err} \leq \err) \wand
      \islazyrand\ lr\ \prop\ \iname\ \gname\wand
      \lazytape\ \lbl\ \None\ \gname \wand
      \upto{\err} \wand \\
      &\pupd[\mask \uplus \{\iname\}]{}\Exists n. \upto{\Err(n)} \sep
      \spac\lazytape\ \lbl\ (\Some n)\ \gname 
    \end{align*}
  \caption{Presampling specification.}
   \end{subfigure}
   ~\\
  \begin{subfigure}[t]{1 \textwidth}
    \begin{align*}
      &\text{R}~n \eqdef{}
        \begin{cases}
          \spac\spac \prop\ n \sep \lazyauth\ n\ \gname \sep \propB_1\ x\ y
          & \text{if }~n = \Some(x, y) \\
          \begin{array}[t]{l}
            \Exists n'. \lazytape\ \lbl\ (\Some n')\ \gname \sep 
            (\lazytape\ \lbl\ \None\ \gname \wand \\
            \qquad \pvs[\top] \prop\ (n',tid) \sep \lazyauth\ (n',tid)\ \gname\sep \propB_2\ n'\ tid)
          \end{array}
          & \text{if }~n = \None
        \end{cases} 
    \end{align*}
    \begin{align*}
      &\spac \hoareV {
        \islazyrand\ lr\ \prop\ \iname\ \gname\sep
        \left(\All n. \prop\ n \wand \lazyauth\ n\ \gname \wand \pupd[\top] R~n\right)
        }{\lazyrandf~lr~\lbl~tid}{(x, y) \ldotp \propB_1\ x\ y \lor \propB_2\ x\ y}[\mask]
    \end{align*}
  \caption{Specification of $\lazyrandf$.}
  \end{subfigure}
  \vspace*{5mm}
  \caption{Excerpt of the specification of the lazy random sampler module.}
  \label{fig:lazy-rand-spec}
\end{figure}

The presampling specification for the lazy random sampler is not too different
from the other previous examples; the main difference is that the abstract tapes
for the module $\lazytape$ stores an option type instead of a list. Since for
each tape, only the first value could ever be relevant in that it is chosen to
be the value stored in the reference, there is no reason to presample more than
one value into a single tape.

Now, let us focus on the more complicated specification for the $\lazyrandf$
function. Firstly, notice that the lazy random sampler predicate $\islazyrand$
takes in an additional predicate $\prop$ as an argument. Intuitively, $\prop$ is
the invariant protected by the lock, and if the reference maps to the value $n$,
it is the case that $\prop\ n$ holds whenever we access the lock and release it.
The precondition of the $\lazyrandf$ function requires two resources. The first
being the abstract predicate $\islazyrand$ and the second being a view shift.
The view shift encodes how the state of the module changes throughout the call.
The view shift starts by assuming that we have $\prop\ n$ and
$\lazyauth\ n\ \gname$ for some $n$, which represents the operation of acquiring
the lock and gaining access to the authoritative state of the lazy random
sampler. We then perform a case distinction on $n$. If it is $\Some (x, y)$, then this
means that the lazy random sampler has already committed to a value, so we
return directly by releasing the lock and establishing some postcondition
$\propB_1\ x\ y$. Otherwise, if it is $\None$, we reach the branch where we have
to do a randomized sampling. Here we are allowed to perform some probabilistic
update operation to provide a non-empty $\lazytape$ (since the view shift is
implemented with a $\pupd[\top]{}$), and it suffices to prove that after reading
that value $n'$ in the tape, we establish the authoritative part of the lazy
random sampler with the reference storing $(n', tid)$ and some postcondition
$\propB_2\ n'\ tid$. If all preconditions hold, then the return value of the
function is some pair $(x,y)$, where either $\propB_1\ x\ y$ or $\propB_2\ x\ y$
holds.

The key ingenuity of the specification of $\lazyrandf$ is that the view shift is
described by the $\pupd{}$ modality, instead of the regular fancy update
modality $\pvs{}$, allowing us to perform presampling on
abstract tapes \emph{within} the function call in addition to outside of it. In
particular, we can choose to perform a presampling action on a tape or not
depending on whether the sampler is storing a $\None$ (it has not been invoked
before) or not, which we know after a case distinction on the value of $n$ after gaining access to the lock.
This flexibility allows us to prove the specification of the
$\lazyrace$, by only performing a \emph{single} presampling within the first invocation
of the function call $\lazyrandf$.

\subsection{Other Case Studies}
\label{sec:other-examples}
Other case studies demonstrating the flexibility of our approach in verifying concurrent randomized
data structures can be found in \appref{sec:app-case-studies}. We  define a Rand module that
captures the operation of sampling from a uniform distribution atomically and we provide three
implementations that satisfy it (similar to the implementations in the randomized counter module
from \cref{sec:three-implementations}). This is the abstract Rand module used to implement the lazy
random sampler in \cref{sec:lazy-rand}. We also implement a concurrent collision-free hash data
structure and show that it meets an amortized specification where the error required for each
operation is amortized across a fixed number of insertions.

\section{Semantic Model and Soundness}
\label{sec:model}

\theaplog is implemented on top of the Iris~\cite{irisjournal} base logic, which in isolation, is simply a higher-order separation logic not tied to any specific programming language. 
In this section, we define the semantic model of \theaplog and explain how to prove the  soundness of the program logic.

\subsection{Model}
\label{subsec:model}
\paragraph{Weakest Precondition}
The \theaplog Hoare triple is defined in terms of a weakest precondition predicate as follows: \[\hoare P e Q \eqdef \always(P \wand \wpre e Q)\]
Expressing Hoare triples in term of a weakest precondition is standard for defining program logics~\cite{irisjournal}, especially for other similar Iris non-probabilistic logics.  
The definition of the weakest precondition is however novel, which we detail below. 
Note that the weakest precondition is defined as a guarded fixed point: the recursive occurrences of the weakest precondition appear under the later modality $\later$ on the last line.
\begin{align*}
  \wpre{\expr_{1}}{\pred} \eqdef{}
  & \All \state_{1}, \err_{1} .
    \stateinterp~\state_{1}~\err_{1} \wand \pvs[\top][\emptyset]\stateStepl{\state_{1}}{\err_{1}}\spac\{ \state_{2},  \err_{2} \ldotp \\
  & \quad \big(\expr_{1} \in \Val \sep\pvs[\emptyset][\top]
    \stateinterp~\state_{2}~\err_{2} \sep \pred~\expr_{1}\big) \lor{} \\
  & \quad
    \big( \expr_{1} \not\in \Val \sep
    \progStepl{(\expr_{1}, \state_{2})}{\err_{2}}
    \spac\{ \expr_{2}, \state_{3}, l, \err_{3} \ldotp  \\
  & \qquad \later \stateStepl{\state_{3}}{\err_{3}}
    \spac\{ \state_{4}, \err_{4} \ldotp 
    \pvs[\emptyset][\top]\stateinterp~\state_{4}~\err_{4} \sep \wpre{\expr_{2}}{\pred} \sep
    \textstyle\Sep_{\expr'\in l} \wpre{\expr'}{\TRUE}
    \} \} \big) \}
\end{align*}
One can intuitively understand $\wpre{\expr}{\pred}$ as a proposition that describes that $e$ is \emph{safe}, meaning it does not get stuck, and that for every possible return value $\val$, the postcondition $\pred~\val$ holds.

We now explain the definition of the weakest precondition in detail.
At the beginning, we assume the ownership of a \emph{state interpretation} $\stateinterp~\state_1~err_1$ for some state $\state_1$ and error value $\err_1$.
This state interpretation $\stateinterp:\State \rightarrow \nnreal \rightarrow \iProp$ gives meaning to the ownership of references $\loc\mapsto\val$, tapes $\progtape{\lbl}{\tapebound}{\tape}$, and error credits $\upto{\err}$.
The resource algebra used to instantiate the state interpretation is standard, and we refer the readers to \citet{eris} for more details.

After that, we perform a view shift through the update modality $\pvs[\top][\emptyset]$, which intuitively means we open all invariants temporarily and that we have access to the resources of all invariants we defined previously.
Following the view shift, we need to prove a \emph{state step precondition} $\stateStepl{\state_1}{\err_1}\{\dots\}$.
The exact definition of the state step precondition is explained later.
For now, we can think of the state step precondition as the modality that allows instantaneous probability-preserving updates supported by the probabilistic update modality $\pupd{\prop}$.
Given state $\state_1$ and error budget $\err_1$, we can perform any number of probability-preserving updates to step to resulting state $\state_2$ and leftover error budget $\err_2$, which must satisfy the rest of the continuation.

The next part of the weakest precondition depends on a case split on the expression $\expr_1$. In the first case, where $\expr_1$ is a value, we do a view shift $\pvs[\emptyset][\top]$ where we re-establish \emph{all} invariants, return the updated state interpretation and show that the return value $\expr_1$ satisfies the postcondition $\pred$. Otherwise, if $\expr_1$ is not a value, we have to prove a {program step precondition}  $\progStepl{(\expr_1,\state_2)}{\err_2}\ \{{\dots}\}$. We later explain the specifics of this precondition modality, but for now, one can loosely understand the connective as somewhat similar to the state step precondition, where we take an actual step on the configuration $(\expr_1, \state_2)$, instead of performing a probabilistic update on $\state_2$. After the single step to resulting expression $\expr_2$, state $\state_3$, forking the list of expressions $l$ with leftover error budget $\err_3$, we prove another state step precondition, which we can ignore here.\footnote{This extra state step precondition is only used to validate certain invariant opening properties not discussed in this paper.} Finally, we re-establish all invariants after the view shift $\pvs[\emptyset][\top]$, return the state interpretation, show that $\wpre{\expr_2}{\pred}$ holds, and $\wpre{\expr'}{\TRUE}$ holds for all $\expr'$ in the forked list $l$.

\begin{figure*}[htb]
  \centering
  \begin{mathpar}
    \inferH{state-step-err-1}
    {1\leq\err}
    { \stateStep{\state}{\err}{\Phi} }
    \and
    \inferH{state-step-ret}
    { \Phi(\state,  \err) }
    { \stateStep{\state}{\err}{\Phi} }
    \and
    \inferH{state-step-continuous}
    { \All \err'.\err<\err' \wand \stateStep{\state}{\err'}{\Phi} }
    { \stateStep{\state}{\err}{\Phi} }
    \and
    \inferH{state-step-exp}
    {
      \expect[\mu]{\Err} \leq \err \\
      \scherasable(\mu, \state_1) \\
      \All \state_2 .
      0<\mu(\state_2)  \wand
      \stateStep{\state_2}{(\Err(\state_2))}{\Phi}
    }
    { \stateStep{\state_1}{\err}{\Phi} }
  \end{mathpar}
  \caption{Inductive Definition of the State Step Precondition $\stateStep{\state}{\err}{\Phi}$.}
  \label{fig:stateStep}
\end{figure*}

\paragraph{State and Program Step Preconditions}
The state step precondition is defined inductively by four inference rules presented in \cref{fig:stateStep}. Firstly, if the error budget $\err$ is larger or equal to $1$, the precondition holds trivially as all sub-distributions have mass smaller or equal to $1$ (\ruleref{state-step-err-1}). If the predicate $\pred$ holds for the current state and error budget, the precondition also holds (\ruleref{state-step-ret}). The third rule \ruleref{state-step-continuous} states that the precondition holds if for error budget \(\err'\) larger than the \(\err\) (in particular, \(\err'\) can be arbitrarily close to \(\err\)), then the precondition does in fact hold for \(\err\) as well. This is the main rule that allows us to create error credits from thin air (\ruleref{pupd-err}), letting us exploit the fact that the real numbers are complete at the level of \theaplog.

Lastly, \ruleref{state-step-exp} is the main interesting rule, which relies on the following auxiliary definition.
\begin{definition}
  A distribution on states $\distr$ is a \defemph{scheduler erasable state update} of $\state \in \State$, written as $\scherasable (\distr,\state)$, if for all schedulers $\sch$, scheduler states $\schstate$, thread pools $\vec e$, and any number of execution steps $n$,
  we have
  \begin{align*}(\distr \mbindi (\Lam \state' . \pexec_{\sch, n} (\schstate, (\vec e, \state')))).\mathsf{tp} \;=\; (\pexec_{\sch, n} (\schstate, (\vec e, \state))).\mathsf{tp}
  \end{align*}
  where we write \(-.\mathsf{tp}\) for the function that projects out the thread pool component from a distribution on configurations.
\end{definition}
A distribution \(\distr\) is thus a scheduler erasable state update of \(\state\) if the probability of executing \(\vec e \) from state \(\state\) to any particular list of threads is the same if we first update the state with respect to \(\distr\) and then execute \(\vec e\). Recall that the operational semantics requires schedulers to be invariant under changes to presampling tapes; such changes thus constitute scheduler erasable state updates.

The \ruleref{state-step-exp} rule then states that
 if we can find a function $\Err:\State \rightarrow [0,1]$ and a distribution $\mu:\Distr{\State}$ such that 
 (1) the expectation of $\Err$ with respect to $\mu$ is at most $\err$, (2) $\mu$ is scheduler erasable with respect to $\state_1$, and (3) for all $\state_2$, the continuation $\stateStep{\state_2}{(\Err(\state_2))}{\pred}$ holds, then $\stateStep{\state_1}{\err}{\pred}$ holds. This is the rule that allows us to do presampling on tapes, since the presampling action is a scheduler erasable operation.

The program step precondition is defined by a single inference rule \ruleref{prog-step-exp}. It is similar to that of \ruleref{state-step-exp}, except that we take exactly one step of the configuration $(\expr_1, \state_1)$. In detail, $\progStep{(\expr_1, \state_1)}{\err}{\pred}$ holds if the configuration $(\expr_1, \state_1)$ is reducible and there exists some function $\Err: \Expr \times \State\times \List(\Expr) \rightarrow [0,1]$ whose expectation with respect to $\stepdistr(\expr_1,\state_1)$ is smaller or equal to $\err$, and for all $(\expr_2, \state_2, l)$, the continuation $\Phi (\expr_2, \state_2, l, \Err(\expr_2, \state_2, l))$ holds.
\begin{mathpar}
  \inferH{prog-step-exp}
  { \red(\expr_1, \state_1) \\
    \expect[\stepdistr (\expr_1, \state_1)]{\Err} \leq \err \\\\
    \rule{0cm}{4mm}
     \All (\expr_2, \sigma_2, l) .0<\stepdistr (\expr_1, \state_1)(\expr_2, \sigma_2, l)\wand
     \Phi (\expr_2, \state_2, l, \Err(\expr_2, \state_2, l)) }
  {\progStep{(\expr_1, \state_1)}{\err}{\Phi}}
\end{mathpar}
\paragraph{Probabilistic Update Modality}

Recall that the  probabilistic update modality $\pupd[\mask_1][\mask_2]P$ depends on two masks $\mask_1$ and $\mask_2$. These extra mask parameters are used to track the opening of invariants and prevent us from opening the same invariant twice (which is unsound). One can understand $\pupd[\mask_1][\mask_2] \prop$ as that we can perform a probability-preserving update to get the resources $\prop$ with the possibility of accessing resources of invariants in the mask $\mask_1$ and reestablishing resources of invariants in the mask $\mask_2$ in the end.
\begin{align*}
  \pupd[\mask_1][\mask_2] P \eqdef{}
  &\All \state_{1}, \err_{1} . 
  \stateinterp~\state_{1}~\err_{1}~ \wand \pvs[\mask_1][\emptyset]\stateStepl{\state_{1}}{\err_{1}}\spac\{ \state_{2},  \err_{2} \ldotp
  \pvs[\emptyset][\mask_2]\stateinterp~\state_{2}~\err_{2} \sep P \}
\end{align*}
The definition of the probabilistic update modality resembles a simplified version of the weakest precondition, where we only perform a single state step. Specifically, $\pupd[\mask_1][\mask_2]\prop$ holds if after assuming some state interpretation $\stateinterp~\state_1~\err_1$, we can open all invariants in $\mask_1$ through the view shift $\pvs[\mask_1][\emptyset]$, and prove a state step precondition with the input parameters $\state_1$ and $\err_1$. Given resulting state $\state_2$ and error budget $\err_2$ after the state step precondition, we re-establish all invariants in the mask $\mask_2$ with the view shift $\pvs[\emptyset][\mask_2]$ and give back the state interpretation and prove the resource $\prop$. 

\subsection{Soundness}
The soundness of \theaplog comes in two flavours, the correctness adequacy theorem \cref{thm:adequacy} and the safety theorem \cref{thm:safety}. We now briefly describe the overall structure proof of the correctness adequacy theorem; the proof of the safety theorem is similar and is omitted.
  
We first prove an intermediate lemma:
\begin{lemma}\label{thm:adequacy_simpl}
  If $\upto{\err}\vdash \wpre{\expr}{\pprop} \sep \Sep_{\expr'\in \vec e'} \wpre{\expr'}{\TRUE}$,
  then for all schedulers $\sch$, states $\state$, and natural numbers $n$,
  $\pr[\execVal_{\sch, n} (\expr\cons \vec e', \state)]{\neg \pprop} \leq \err$. %
\end{lemma}

This lemma is proven by induction on $n$ and structural induction on the state step precondition fixed point.
For each step, we unfold the definition of $\execVal$ to determine which thread the scheduler chooses to step next. We unfold the definition of the corresponding weakest precondition proposition (the one that matches the thread chosen to step), and show that the $\stateStepp$  and $\progStepp$ modalities satisfies monadic composition, allowing us to compose the errors.

By taking $\vec e'$ to be the empty list of threads in \cref{thm:adequacy_simpl}, and taking the limit of $n$, we can then show $\pr[\limexecVal_\sch \expr]{\neg \pprop} \leq \err$, which is the goal of the adequacy theorem.

\section{Related Work}
\label{sec:related-work}
\paragraph{Approximate Reasoning}
There are various approaches for tracking error probabilities in probabilistic programs. Approximate Hoare logic~\cite{union_bound_logic} uses a grading on Hoare triples to approximate error probabilities. Expectation-based logics such as that of \citet{qsl,ppt} are defined with a weakest-precondition-style quantitative predicate transformer that computes the expected value of a program's postcondition, which can be used to derive approximate correctness bounds. Compared to our work, these logics are usually restricted to sequential, first-order imperative programs.  Our method of using error credits to track error bounds is first used in Eris~\cite{eris} to prove error bounds of sequential higher-order probabilistic programs.

Various other logics also considered reasoning about approximate correctness in the relational setting. apRHL~\cite{aprhl, aprhl2} relates the probability distribution of two programs through approximate probabilistic couplings, which can then be used to prove differential privacy. Inspired by Eris, error credits are used in Approxis~\cite{approxis} to prove approximate equivalences of higher-order programs. 
\paragraph{Concurrent Probabilistic Program Logics}

One of the first program logics developed for concurrent probabilistic programs is the  probabilistic rely-guarantee calculus~\cite{prob-rely-guarantee} (that extends the rely-guarantee logic~\cite{rely-guarantee}) that verifies the quantitative correctness of a probabilistic concurrent programs without local state.
Later, Concurrent Quantitative Separation Logic~\cite{cqsl} extends Quantitative Separation Logic~\cite{qsl} to reason about the lower bounds of probability to realize the postcondition of concurrent, heap-manipulating, randomized imperative programs. Compared to our work, it cannot establish strict error bounds that arise between the interleaving of threads (see the $\contwodie$ example in \cref{sec:tech-overview}) and cannot reason about programs in a (procedure-)modular way.

Polaris~\cite{polaris} is a logic for establishing refinements between concurrent probabilistic programs and a monadic representation via probabilistic couplings inspired by pRHL~\cite{uniform-independence,prhl-proof,coupling-proof}. The simpler monadic model can then be studied to derive properties of the original programs, such as bounds on its expected value. The language considered by \theaplog is inspired by that of Polaris; the syntax is the same, but in \theaplog, we allow schedulers to be probabilistic as well. 
Compared to \theaplog, Polaris is not as modular in the sense that it does not demonstrate how to compose refinements of different data structures. It also does not develop an approach for reasoning about logical atomicity.

\citet{lohse2024iris} develop ExpIris, a variant of Iris that supports establishing bounds on the expected cost of concurrent higher-order programs with mutable state.
In ExpIris, an upper bound budget on the number of steps a program can take is written as an additional parameter of weakest preconditions, called a \emph{potential}.
On randomized steps, this potential can be updated in an expectation-preserving way, similar to \ruleref{ht-rand-exp}.
However, because potentials are a parameter of the weakest precondition, instead of a separation logic resource like error credits, it is not possible to share them in an invariant, as we saw was necessary for obtaining tight analyzes in \cref{sec:tech-overview}. ExpIris also does not provide any facilities to encode the notion of randomized logical atomicity, which we show is essential to reason about concurrent programs modularly. 
Although ExpIris provides rules for reasoning about concurrency, there are no case studies provided that utilize concurrent constructs (\eg{}, the $\Fork$ construct). 

Recently, Probabilistic Concurrent Outcome Logic~\cite{pcol} extends Demonic Outcome Logic~\cite{dol} to reason about the distributions of outcomes from concurrent probabilistic programs. Although this logic is able to prove other probabilistic properties beyond the scope of \theaplog, such as independence and conditioning, the programs considered are restricted to those without dynamically allocated state or higher-order functions, and the logic does not support defining ghost state.

\paragraph{Internalization of Linearizability}
There is a long line of research on internalizing linearizability as a reasoning principle within concurrent program logic specifications. \citet{jacobs-piessens} first extended the resource-invariants-based method from Owicki and Gries~\cite{resource-invariants}  allowing users to parameterize the specification of concurrent functions with ghost code. Later, \citet{hocap} further extended their idea and proposed a new style of specification using higher-order concurrent abstract predicates (HOCAP), building on top of CAP~\cite{cap}. \citet{tada} introduced a different logic called TaDa, which proposed the use of atomic triples to capture logical atomicity of programs. There has also been much research in encoding logically atomic specifications within the Iris separation logic~\cite{iris, iris-prophecy}. 
In \theaplog, we take inspiration from these logics, especially HOCAP, to capture randomized logical atomicity within \emph{probabilistic} concurrent programs. 

\section{Conclusion}
\label{sec:conclusion}
We presented \theaplog, the first concurrent and probabilistic higher-order separation logic for error bound reasoning. \theaplog captures randomized logical atomicity through the novel probabilistic update modality, enabling modular verification of concurrent programs that is out-of-scope for previous techniques. We demonstrated the flexibility of \theaplog by verifying various examples modularly, most of which involve local state and intricate reasoning over randomness that arise from concurrency. 

There are various directions for extending \theaplog. Firstly, we would like to extend \theaplog to enable verifying strict error bounds of concurrent probabilistic programs under restricted schedulers, such as those that cannot view the configuration of the program. It is also interesting to explore whether ideas from Approxis~\cite{approxis} can be used to extend \theaplog into the relational setting to establish approximate bounds between concurrent probabilistic programs. Lastly, we would like to consider integrating cost credits from Tachis~\cite{tachis} into \theaplog to reason about both the expected work and span time costs of concurrent probabilistic programs.

\section*{Data Availability Statement}
The Rocq formalization accompanying this work is available on Zenodo \cite{zenodo:coneris}.

\begin{acks}
  The first author would like to thank Amin Timany for enlightening discussions regarding HOCAP-style specifications.
  The authors also thank Fran\c{c}ois Pottier for finding an error in an earlier description of the example in \cref{sec:tech-overview}.
  This work was supported in part by the \grantsponsor{NSF}{National Science Foundation}{}, grant no.~\grantnum{NSF}{2338317}, the \grantsponsor{Carlsberg Foundation}{Carlsberg Foundation}{}, grant no.~\grantnum{Carlsberg Foundation}{CF23-0791}, a \grantsponsor{Villum}{Villum}{} Investigator grant, no. \grantnum{Villum}{25804}, Center for Basic Research in Program Verification (CPV), from the VILLUM Foundation, and the European Union (\grantsponsor{ERC}{ERC}{}, CHORDS, \grantnum{ERC}{101096090}).
  Views and opinions expressed are however those of the author(s) only and do not necessarily reflect those of the European Union or the European Research Council.
  Neither the European Union nor the granting authority can be held responsible for them.
\end{acks}

\bibliography{refs}
\ifbool{fullversion}{
 \pagebreak
 \appendix
 
 \section{Modular Proof of $\contwodie$}
\label{sec:app-proof-counter-client}
In this section, we show in more detail how to prove $\contwodie$ with the HOCAP-style specifications
of the randomized counter module (see \cref{fig:rand-counter-spec-with-tapes}).

Before we proceed, we present a selection of side conditions of the abstract predicates   in \cref{fig:side-conditions-on-counter} which we previously omitted in \cref{fig:rand-counter-spec-with-tapes}. The first side condition expresses
that the counter representation predicate is persistent, which means that it is
duplicable so that clients can share it among several threads.
We then have a series of side conditions regarding the $\cauth$ and $\cfrag$ abstract predicates, which are used to keep track of the abstract state of the counter. The first condition states that $\cfrag$ abstract predicates can be combined by adding their arguments together. %
The next condition states that if we hold both the $\cauth$ and $\cfrag$ resource and the fraction of the $\cfrag$ is exactly $1$, the values from both predicates agree.  The  last side condition describes how we can update the abstract state of a counter: if we have a $\cauth$ and a $\cfrag$ predicate with the same ghost name, we can update the predicates by incrementing the values of both by a constant $x$.
\begin{figure}[htbp]
  \centering
  \begin{align*}
     C~\iname~\gname~c& \wand \always C~\iname~\gname~c  \\
    \cfrag\ \gname\ f\ z& \wand \cfrag\ \gname\ f'\ z' \wand \cfrag\ \gname\ (f+f')\ (z+z') \\
    \cauth\ \gname\ z& \wand \cfrag\ \gname\ 1\ z' \wand z'= z \\
    \cauth\ \gname\ z &\wand \cfrag\ \gname\ f\ z' \wand\pvs \cauth\ \gname\ (z+x) \wand \cfrag\ \gname\ f\ (z'+x) 
  \end{align*}                                
 \caption{Selection of Side Conditions on Abstract Predicates}
  \label{fig:side-conditions-on-counter}
\end{figure}

Recall that since the new specification of the randomized concurrent counter utilizes tapes,
the $\contwodie$ client is annotated to use the abstract tapes.
     \begin{align*}
       \contwodie \eqdef{}&
       \Let c = \createcounter\TT in  \\
       &\big(\Parallel{
         \begin{array}{c}
           \Let \lbl=\createtape \TT in\\
           \incrcounter\ c\ \lbl\\
         \end{array}
       }{\begin{array}{c}
           \Let \lbl=\createtape \TT in\\
           \incrcounter\ c\ \lbl\\
       \end{array}}\big); \\
       &\readcounter\ c
     \end{align*}
We now prove that the return value is $0$ with a probability of $1/16$,
with the  \theaplog Hoare triple: $\hoare{\upto{1/16}}{\contwodie}{v.v>0}$.

We first consider the invariant used to track the change in shared state
during the parallel composition. We use two states $S_0$ and $S_1(n)$ (of some
inductive type $T$) to track the state of the threads, with $S_0$ representing
the state where the thread has not sampled a value yet and $S_1(n)$ representing it
sampled $n$. Note that we do not need an additional state to track whether the
sampled value has been added into the counter, because that can be tracked by
the resource $\cfrag$. We use the invariant $I$ shown below to capture the shared state of
the two threads. 
Notice that the invariant $I$ makes use of the \emph{exclusive-authoritative} ghost resource algebra, which consists of the \emph{authoritative} part $\authfull x$ and the \emph{fragment} part $\authfrag x$. We omit the definition and properties of this resource and we refer readers to \citet{irisjournal} for more information.
\begin{align*}
  \sampled\ s \eqdef{}& \langkw{match}\ s\ \langkw{with} \ \ S_0 \Ra \None\  |\ S_1 (n) \Ra \Some n\ \langkw{end}\\
 \onepositive\ s_1\  s_2\eqdef{}& \Exists n . n>0 \land (\sampled\ s_1 = \Some n \lor \sampled\ s_2 = \Some n)\\
  I(\gname_1, \gname_2)\eqdef{} &\Exists (s_1\ s_2: T)  . \ownGhost{\gname_1}{\authfull s_1} \sep \ownGhost{\gname_2}{\authfull s_2} \sep\\
  & \spac \textlog{if}\ \onepositive\ s_1\ s_2\ \textlog{then}\ \upto{0} \\
  & \spac \spac \textlog{else}\ \upto{4^{(\booltonat (\sampled\ s_1=\Some 0)+\booltonat (\sampled\ n_2=\Some 0)-2)}}
\end{align*}
We now show how to prove the specification of $\contwodie$ using the invariant previously defined.
After stepping through the code up until the parallel composition component, and
allocating the necessary resources and invariant, we arrive at the following proof
obligation:
\begin{align*}
  \hoare{
    \begin{array}{c}
      C~\iname~\gname~c \sep \cfrag\ \gname\ 1\ 0 \sep \knowInv{\iname'}{I(\gname_1, \gname_2)} \sep\\
      \ownGhost{\gname_1}{\authfrag S_0}\sep \ownGhost{\gname_2}{\authfrag S_0}
  \end{array}}{
      \begin{array}{l}
       \Parallel{\langkw{let}\dots}{\langkw{let}\dots}; \\
       \readcounter\ c
      \end{array}
  }{v\ldotp v>0}
\end{align*}
We can apply the side condition of $\cfrag$ to split it between the two threads
and apply the rule for parallel composition which leaves us with the following three
obligations:
\begin{align}
    &\hoare{C~\iname~\gname~c \sep  \knowInv{\iname'}{I(\gname_1, \gname_2)} \sep \cfrag\ \gname\ 0.5\ 0 \sep\ownGhost{\gname_1}{\authfrag S_0}}{
      \langkw{let}\dots
    }{\Exists n. \cfrag\ \gname\ 0.5\ n \sep \ownGhost{\gname_1}{\authfrag S_1(n)}} \label{eq:contwodie1}\\
    &\hoare{C~\iname~\gname~c  \sep \knowInv{\iname'}{I(\gname_1, \gname_2)}\sep \cfrag\ \gname\ 0.5\ 0 \sep \ownGhost{\gname_2}{\authfrag S_0}}{
      \langkw{let}\dots
    }{\Exists n. \cfrag\ \gname\ 0.5\ n \sep \ownGhost{\gname_1}{\authfrag S_2(n)}} \label{eq:contwodie2}\\
  &\hoare{\begin{array}{c}
      C~\iname~\gname~c \sep \cfrag\ \gname\ 1\ (n_1+n_2) \sep \knowInv{\iname'}{I(\gname_1, \gname_2)} \sep \\
      \ownGhost{\gname_1}{\authfrag S_1(n_1)}\sep \ownGhost{\gname_2}{\authfrag S_1(n_2)}
      \end{array}
  }{
      \readcounter\ c
  }{v\ldotp 0<v} \label{eq:contwodie3}
\end{align}
Let us first focus on \cref{eq:contwodie1}.
We first apply the specification for $\createtape$ to create an empty tape resource and we arrive at the following obligation.
\begin{mathpar}
  \hoare{
    \begin{array}{c}
      C~\iname~\gname~c \sep  \knowInv{\iname'}{I(\gname_1, \gname_2)} \sep \cfrag\ \gname\ 0.5\ 0 \sep\\
      \ownGhost{\gname_1}{\authfrag S_0} \sep \ctape\ \lbl\ \nil
  \end{array}}{
      \incrcounter\ c\ \lbl
    }{\Exists n. \cfrag\ \gname\ 0.5\ n \sep \ownGhost{\gname_1}{\authfrag S_1(n)}}
\end{mathpar}
Now that we have a $\ctape$ predicate on our hands, we can presample a
value onto it so that it can be used for the $\incrcounter$ method later.
Specifically, we perform the following probabilistic update:
\[
  \knowInv{\iname'}{I(\gname_1,\gname_2)}\sep \ownGhost{\gname_1}{\authfrag S_0} \sep \ctape\ \lbl\ \nil \wand  \spac\pupd[\top]\Exists n. \ownGhost{\gname_1}{\authfrag S_1 (n)} \sep \ctape\ \lbl\ [n]
\]
This probabilistic update proposition is proven by first applying the probabilistic update modality version of \ruleref{inv-open} where we  access the resources within the invariant, and subsequently updating the authoritative resource pairs
from the state $S_0$ to $S_1(n)$ to track the value $n$ presampled onto the tape.
After this probabilistic update, we are left with the following obligation:
\begin{mathpar}
  \hoare{
    \begin{array}{c}
      C~\iname~\gname~c \sep  \knowInv{\iname'}{I(\gname_1, \gname_2)} \sep \cfrag\ \gname\ 0.5\ 0 \sep\\
      \ownGhost{\gname_1}{\authfrag S_1(n)} \sep \ctape\ \lbl\ [n]
  \end{array}}{
      \incrcounter\ c\ \lbl
    }{\Exists n. \cfrag\ \gname\ 0.5\ n \sep \ownGhost{\gname_1}{\authfrag S_1(n)}}
\end{mathpar}
The rest of the proof then follows almost directly by applying the new specification for $\incrcounter$ and choosing $\propB\ z \eqdef{} \cfrag\ \gname\ 0.5\ n$. The second obligation (\cref{eq:contwodie2}), representing the behavior of the second thread, is proven in an almost identical fashion.

Let us now focus on the last obligation (\cref{eq:contwodie3}). To prove that the return value is positive, we apply the specification of $\readcounter$, choosing $\propB\ v \eqdef{}v>0$, leaving us with the following view shift obligation for the precondition:
\begin{align*}
  & \knowInv{\iname'}{I(\gname_1, \gname_2)} \sep \cfrag\ \gname\ 1\ (n_1+n_2) \sep \ownGhost{\gname_1}{\authfrag S_1(n_1)}\sep \ownGhost{\gname_2}{\authfrag S_1(n_2)} \wand \\
  &\spac (\All z. \cauth\ \gname\ z \wand \pvs[\top\setminus\iname] \cauth\ \gname\ z \sep z>0)
\end{align*}
We do a case analysis on the values of $n_1$ and $n_2$. If they are both $0$, we
can open the invariant $I(\gname_1, \gname_2)$ to access a $\upto{1}$ error
credit to derive a contradiction with \ruleref{err-1}. Otherwise, using the rules for
$\cauth$ and $\cfrag$, we can show that the values in the $\cauth$ and
$\cfrag$ predicates coincide, i.e. they are both $n_1+n_2$ and must be positive, which
completes the proof.

 \section{HOCAP-style Specification with Error Redistribution}
\label{sec:hocap-spec-with-error}
In \cref{sec:tech-overview-modularity}, we presented a HOCAP-style specification that does not expose presampling tapes as an abstract predicate (see \cref{fig:rand-counter-spec-with-errors}). Although in \cref{sec:spec-errors-tapes-proof} we showed that it is less general than the specification with $\ctape$ shown in \cref{fig:rand-counter-spec-with-tapes}, in this section, we briefly explain how to use the specification to prove clients of the module and how to show various implementations meet the specification.
\subsection{Implementation}
We first show three possible implementations of the module that mirror those shown in \cref{sec:three-implementations}. 
\begin{figure}[htbp]
  \centering
  \begin{align*}  
    \incrcounter_1 \eqdef{} & \Lam l. \Faa l\ (\Rand 3) \\
    \incrcounter_2 \eqdef{} & \Lam l. \Let \lbl = \AllocTape 1 in \\
    & \spac \Faa l\ (\Rand \lbl\ 1*2 + \Rand \lbl\ 1)  \\
    \incrcounter_3 \eqdef{} & \Lam l. \Let \lbl = \AllocTape 4 in \\
    & \spac (\Rec f \lbl=\\
                            &\spac\spac\DumbLet x = \Rand\lbl\ 4 \In \\
                            & \spac \spac \spac \If x<4 then \Faa l\ x \Else f\ \lbl)\ \lbl  
  \end{align*}                                
  \caption{Three Implementations of Increment}
  \label{fig:incr-implementations}
\end{figure}
For $I_1$, we do not need to allocate any tapes and we sample from $\Rand 3$ directly.
However, note that for $I_2$ and $I_3$, we have to create a tape internally and sample from it. This is because the randomization within the two implementations occurs over various steps even if it acts ``logically atomic''. By adding extra ghost code that utilizes tapes, we are able to reason about the randomness asynchronously, which we demonstrate in later subsections.

\subsection{Verification of Client of HOCAP-style Specification with Error Redistribution}
We now verify the $\contwodie$ example in \cref{sec:spec-errors-tapes-proof} with the specification from \cref{fig:rand-counter-spec-with-errors}. Since tapes are not exposed in this specification, $\contwodie$ is written without explicit allocation of $\ctape$.
\begin{align*} 
  \contwodie \eqdef{} &\Let c = \createcounter\TT in  \\
  & (\Parallel{\incrcounter\ c}{\incrcounter\ c}); \\
  & \readcounter\ c
\end{align*}

As before, we want to verify that the final read value is positive, with error probability $1/16$,
which we show with the following \theaplog Hoare triple.
\[\hoare{\upto{1/16}}{\contwodie}{v.v>0}\]
In fact the proof works almost identically to that presented in \cref{sec:app-proof-counter-client}.
For example, the states and invariants used to track the shared state of the two parallel threads are
identical to the ones used before.

We begin by stepping through the code up until the parallel composition component, and after
allocating the necessary resources and invariant, we arrive at the following proof
obligation:
\begin{align*}
    \hoareV{C~\iname~\gname~c \sep \cfrag\ \gname\ 1\ 0 \sep \knowInv{\iname'}{I(\gname_1, \gname_2)} \sep \ownGhost{\gname_1}{\authfrag S_0}\sep \ownGhost{\gname_2}{\authfrag S_0} }{
      \begin{array}{l}
       \Parallel{\incrcounter\ c}{\incrcounter\ c}; \\
       \readcounter\ c
      \end{array}
  }{v\ldotp v>0}
\end{align*}
We can apply the side condition of $\cfrag$ to split it between the two threads
and apply the rule for parallel composition, leaving us with the following three
obligations:
\begin{align}
    &\hoare{C~\iname~\gname~c \sep  \knowInv{\iname'}{I(\gname_1, \gname_2)} \sep \cfrag\ \gname\ 0.5\ 0 \sep\ownGhost{\gname_1}{\authfrag S_0}}{
      \incrcounter\ c
    }{\Exists n. \cfrag\ \gname\ 0.5\ n \sep \ownGhost{\gname_1}{\authfrag S_1(n)}} \label{eq:contwodie4}\\
    &\hoare{C~\iname~\gname~c  \sep \knowInv{\iname'}{I(\gname_1, \gname_2)}\sep \cfrag\ \gname\ 0.5\ 0 \sep \ownGhost{\gname_2}{\authfrag S_0}}{
      \incrcounter\ c
    }{\Exists n. \cfrag\ \gname\ 0.5\ n \sep \ownGhost{\gname_1}{\authfrag S_2(n)}} \label{eq:contwodie5}\\
  &\hoare{C~\iname~\gname~c \sep \cfrag\ \gname\ 1\ (n_1+n_2) \sep \knowInv{\iname'}{I(\gname_1, \gname_2)} \sep \ownGhost{\gname_1}{\authfrag S_1(n_1)}\sep \ownGhost{\gname_2}{\authfrag S_1(n_2)}}{
      \readcounter\ c
  }{v\ldotp 0<v} \label{eq:contwodie6}
\end{align}
We focus only on the first obligation; the second obligation follows similarly and the third obligation is similar to the proof of \cref{eq:contwodie3} in \cref{sec:app-proof-counter-client}.
From \cref{eq:contwodie4}, we apply the
specification of $\incrcounter$ directly, choosing
$\propB\ \err\ \Err\ n\ z\eqdef{} \cfrag\ \gname\ 0.5\ n \sep
\ownGhost{\gname_1}{\authfrag S_1(n)}$. It then suffices to prove the following
view shift for the precondition of the specification:
\begin{align*}
  &\knowInv{\iname'}{I(\gname_1, \gname_2)}\sep \cfrag\ \gname\ 0.5\ 0 \sep \ownGhost{\gname_1}{\authfrag S_0} \wand \\
  &\spac \pvs[\mask][\emptyset] \Exists \err\ \Err. \upto{\err} \sep (\expect[\unifd{3}]{\Err} \leq \err) \sep \\
    &  \spac(\All x. 0\leq x <4  \wand \upto{\Err(x)} \wand \pvs[\emptyset][\mask] \\
    &\spac \spac(\All z. \cauth\ \gname\ z \wand \pvs[\mask] \cauth\ \gname\ (z+x) \sep \cfrag\ \gname\ 0.5\ x \sep \ownGhost{\gname_1}{\authfrag S_1(x)})
    )    
\end{align*}
We first open our invariant $I(\gname_1, \gname_2)$ while stripping away the $\pvs[\mask][\emptyset]$ mask, which allows us to access the error credit stored in the invariant. After choosing the right $\Err$ based on a case analysis on the state of the right thread (which we omit for brevity), we update the authoritative resource pairs from $\ownGhost{\gname_1}{\authfull S_0}\sep\ownGhost{\gname_1}{\authfrag S_0}$ to $\ownGhost{\gname_1}{\authfull S_1(x)}\sep\ownGhost{\gname_1}{\authfrag S_1(x)}$ and re-establish the invariant $I$ while removing the $\pvs[\emptyset][\mask]$ mask, leaving us with the following state:
\begin{align*}
  &\knowInv{\iname'}{I(\gname_1, \gname_2)}\sep \cfrag\ \gname\ 0.5\ 0 \sep \ownGhost{\gname_1}{\authfrag S_1(x)} \wand \\
  & \spac \cauth\ \gname\ z \wand \pvs[\mask] \cauth\ \gname\ (z+x) \sep \cfrag\ \gname\ 0.5\ x \sep \ownGhost{\gname_1}{\authfrag S_1(x)}    
\end{align*}
After incrementing both the $\cauth$ and $\cfrag$ components by exactly $x$ through the $\pvs[\mask]$ mask (which is possible by the side conditions of $\cauth$ and $\cfrag$), we can then directly establish the final goal. %

\subsection{Proving that $I_{1}$, $I_{2}$, and $I_{3}$ Satisfy the HOCAP-style Specification with Error Redistribution}
We now briefly describe how each of the three randomized counter implementations meets the
specification with error redistribution in
\Cref{fig:rand-counter-spec-with-errors}.
The concrete definitions for the abstract predicates are actually identical to those
used in the proof of \cref{sec:prove-implementation-counter}. For example, the counter predicate
is still defined as
\[\knowInv{\iname}{\Exists (l:\Loc) (n:\tnat). c=l \sep l\mapsto n \sep \cauth\
  \gname\ n}\]
It then suffices to
show that the functions $\createcounter, \incrcounter$, and $\readcounter$
satisfy the HOCAP-style specification. We focus on the $\incrcounter$
specification since it is the most complicated; the other two functions can be
verified in a similar, if not easier, fashion.

For $I_{1}$, after symbolically stepping through the program, we are left with
the following obligation:
\begin{mathpar}
\hoareVH {
    \begin{array}{l}
      C~\iname~\gname~c\sep (\pvs[\mask][\emptyset] \Exists \err\ \Err.
      \spac \upto{\err} \sep (\expect[\unifd{3}]{\Err} \leq \err) \sep \\
      \spac(\All x. 0\leq x <4  \wand \upto{\Err(x)} \wand \pvs[\emptyset][\mask] \\
    \spac \spac(\All z. \cauth\ \gname\ z \wand \pvs[\mask] \cauth\ \gname\ (z+x) \sep Q\ \err\ \Err\ x\ z)
    )
      )
      \end{array}
}{\Faa c\ (\Rand 3)}{z. \Exists \err\ \Err\ x. Q\ \err\ \Err\ x\ z}[\mask \uplus \{\iname\}]
\end{mathpar}
We first open the $\pvs[\mask][\emptyset]$ mask around the atomic $\Rand 3$
operation. After opening the first view shift, we are given some $\upto{\err}$
and some $\Err$ such that the expected sum of $\Err$ is smaller than $\err$.
We then apply \ruleref{ht-rand-exp} to distribute the errors across the various
results and close the $\pvs[\emptyset][\mask]$ mask, leaving us with the
following obligation:
\begin{mathpar}
\hoareVH {
    \begin{array}{l}
      C~\iname~\gname~c\sep (\All z. \cauth\ \gname\ z \wand \pvs[\mask] \cauth\ \gname\ (z+x) \sep Q\ \err\ \Err\ x\ z)
      \end{array}
}{\Faa c\ x}{z. \Exists \err\ \Err\ x. Q\ \err\ \Err\ x\ z}[\mask \uplus \{\iname\}]
\end{mathpar}
Since the $\langkw{faa}$ operation is atomic, we can open the invariant $C$
around the expression, resulting in this goal:
\begin{mathpar}
\hoareVH {
    \begin{array}{l}
       l\mapsto n \sep \cauth\ \gname\ n \sep (\All z. \dots )
      \end{array}
}{\Faa l\ x}{z. \Exists (n:\tnat). l\mapsto n \sep \cauth\ \gname\ n \sep \Exists \err\ \Err\ x. Q\ \err\ \Err\ x\ z}[\mask]
\end{mathpar}
The rest of the proof follows nicely from the proof rule for the $\langkw{faa}$ operation, completing the proof.

For $I_{2}$, we similarly step through the program,
where we additionally allocate a tape $\lbl$, and thus we arrive at the following goal:
\begin{mathpar}
\hoareVH {
    \begin{array}{l}
      C~\iname~\gname~c\sep \mhl{\progtape{\lbl}{1}{\nil} \sep} (\pvs[\mask][\emptyset] \Exists \err\ \Err.\\
      \spac \upto{\err} \sep (\expect[\unifd{3}]{\Err} \leq \err) \sep \\
      \spac(\All x. 0\leq x <4  \wand \upto{\Err(x)} \wand \pvs[\emptyset][\mask] \\
    \spac \spac(\All z. \cauth\ \gname\ z \wand \pvs[\mask] \cauth\ \gname\ (z+x) \sep Q\ \err\ \Err\ x\ z)
    )
      )
      \end{array}
}{\Faa l\ (\Rand \lbl\ 1*2 + \Rand \lbl\ 1)}{z. \Exists \err\ \Err\ x. Q\ \err\ \Err\ x\ z}[\mask \uplus \{\iname\}]
\end{mathpar}
From here, we directly open the $\pvs[\mask][\emptyset]$ and access the
$\upto{\err}$ error credit. Then, unlike what we did for $I_1$, here we perform
a \emph{probabilistic update}  where we
presample two values $v_1, v_2$ onto the tape $\lbl$, and we distribute
$\upto{v_1*2 + v_2}$ for each branch, i.e., we update the resources via the
following lemma (in this instance, $\tape$ is instantiated to be the empty tape
list $\nil$). This follows from \cref{eq:presample-spec2} which we proved previously.

After closing the $\pvs[\emptyset][\mask]$ mask, we are left with the following obligation:
\begin{mathpar}
\hoareVH {
    \begin{array}{l}
      C~\iname~\gname~c\sep \progtape{\lbl}{1}{[v_1,v_2]} \sep \\
    \spac(\All z. \cauth\ \gname\ z \wand \pvs[\mask] \cauth\ \gname\ (z+v_1*2+v_2) \sep Q\ \err\ \Err\ (v_1*2+v_2)\ z)  
      \end{array}
}{\Faa l\ (\Rand \lbl\ 1*2 + \Rand \lbl\ 1)}{z. \Exists \err\ \Err\ x. Q\ \err\ \Err\ x\ z}[\mask \uplus \{\iname\}]
\end{mathpar}
We can then read the values of the tape directly for both samples:
\begin{mathpar}
\hoareVH {
    \begin{array}{l}
      C~\iname~\gname~c\sep \progtape{\lbl}{1}{\nil} \sep \\
    \spac(\All z. \cauth\ \gname\ z \wand \pvs[\mask] \cauth\ \gname\ (z+v_1*2+v_2) \sep Q\ \err\ \Err\ (v_1*2+v_2)\ z)  
      \end{array}
}{\Faa l\ (v_1*2 + v_2)}{z. \Exists \err\ \Err\ x. Q\ \err\ \Err\ x\ z}[\mask \uplus \{\iname\}]
\end{mathpar}
From here, the fetch-and-atomic-add step is similar to that for the $I_1$ implementation.

The proof of $I_3$ is very similar to that of $I_2$, except for the  presampling step after allocating the tape resource. In particular we want to show that we can repeatedly presample enough values into the tape such that the last element is smaller than $4$ and all values beforehand are $4$, while distributing the error credit according to the final value. 
Here we use \cref{eq:presample-spec3} proved previously to do so.

After performing the probabilistic update on the tape (such that it contains an ``accepted'' value at the end), we can then step through the rest of the program, looping repeatedly until we reach the final ``accepted'' value and establish the postcondition.

 \section{Other Case Studies}
\label{sec:app-case-studies}
\subsection{Rand Module}
\label{sec:rand-module}
For the random counter module introduced previously in
\cref{sec:tech-overview-modularity}, we identified three distinct
implementations (\cref{sec:three-implementations}) that sample randomness from a
uniform distribution for the $\incrcounter$ operation, e.g.~we can directly call
a single $\Rand$ ($I_1$), chain various $\Rand$s together ($I_2$), or use a
rejection sampler method where we repeatedly sample until we obtain a desirable
value ($I_3$). We now define a general interface, which we refer to as the
\emph{Rand module}, that captures what it means to sample from a uniform
distribution atomically, and show that several implementations satisfy it. In
later case studies, we use this interface to verify larger programs, to
highlight the usability of this module and to demonstrate modular reasoning.

The Rand module is parameterized by a natural number $\tapebound$, which is the
range of the uniform distribution (we are sampling uniformly from
$\{0,\dots,\tapebound\}$). The interface exposes two functions, $\randallocate$
and $\randf$, which creates a tape and samples from it, respectively. It also
describes various abstract predicates, their side conditions, and specifications
of the functions, most notably the specification that allows clients to
presample into the abstract tape $\randtape$, and reading from it with $\randf$,
which we present in \cref{fig:rand-module-spec}.
\begin{figure}
  \begin{subfigure}[t]{0.43\textwidth}
    \label{fig:rand-module1}
\begin{align*}
             & \begin{pmatrix*}[l]
        \All \err\ \Err\ \iname\ \mask\ \lbl\ \gname\ \tape. \\
        \spac(\expect[\unifd{\tapebound}]{\Err} \leq \err) \wand\\
    \spac\iname\in\mask\wand\\
    \spac \isrand\ \iname\ \gname\wand\\
    \spac \randtape\ \lbl\ \tape\ \gname \wand \\
    \spac \upto{\err} \wand \\
    \spac \pupd[\mask]{\Exists n. \upto{\Err(n)} \sep \randtape\ \lbl\ (\tape\lapp [n])\ \gname}
               \end{pmatrix*}
  \end{align*}
  \caption{Presampling specification }
   \end{subfigure}
   ~
  \begin{subfigure}[t]{0.43\textwidth}
    \label{fig:rand-module2}
\begin{align*}
   & \All \iname\ \gname\ \lbl\ n\ \tape\ \mask.\\
  &\spac \hoareVH {
    \begin{array}{l}
      \iname\in\mask\sep\isrand\ \iname\ \gname\sep {\ctape\ \lbl\ (n\cons\tape)}
      \end{array}
  }{\randf\ \lbl}{z. {z=n\sep\randtape\ \lbl\ \tape}}[\mask]
\end{align*}
  \caption{$\randf$ specification }
\end{subfigure}\vspace*{5mm}
  \caption{Selection of Specification of  Rand Module}
  \label{fig:rand-module-spec}
\end{figure}

The first condition states that when given the $\isrand$ invariant, a
$\randtape$ abstract predicate, and some error credits, we can append a value at
the end of the tape and split the errors in an expectation-preserving way,
similar to the presampling specification presented in the random counter module.
The second condition states that given the $\isrand$ invariant and a non-empty
tape, we can run $\randf$ on the tape to deterministically pop the tape and
return its first element.

By choosing concrete definitions for the abstract predicates of this module,
one can show that various implementations of random samplers satisfy this Rand
module specification;  e.g. we proved that a rejection sampler meets the
specification of the Rand module (the proof is similar to showing that implementation $I_3$
of the randomized counter module meets its specification).

\subsection{Concurrent Amortized Collision-free Hash}
\label{sec:amortized-concurrent-hash}
When proving the correctness of randomized data structures, it is useful to assume that
a hash function is collision-free, in that different input keys for the function return different hash values. In reality, collisions might occur but with very low probabilities.

Here, we first consider a concurrent collision-free hash, one that can be shared among
many threads, and each thread pays error credits to avoid collisions for every presampling action.
Because we want to be able to use the hash in a concurrent context, the specification of the hash is
written in a  HOCAP-style with presampling tapes exposed to allow modular reasoning.
We implement a concurrent model of the idealized hash function under the uniform
hash assumption~\cite{uniform-hash-assumption}. The assumption states that the
hash function $\hashf$ mapping sets of keys $K$ to hash values $V$ is a random
oracle, in that for each key $k\in K$, the hash value $h(K)$ is sampled
uniformly from $V$ independently of all other keys. We implement this model as a
tuple containing a lock and a mutable map $lm$, choosing $K$ and $V$ to be $\{0,\dots,\tapebound\}$.
The main hash function $\hashf$ is shown below. The lock is acquired and
released around the body of the hash function to ensure that  at most one thread is
changing the state of the mutable map. In the critical section, if the key
$k$ has been hashed before, we directly return $lm(k)$. Otherwise, we sample a
fresh value uniformly from $V$ with the $\randf$ function defined in
\cref{sec:rand-module}, read a value from the tape $\lbl$, store it in $lm(k)$, 
and return it at the end.
\begin{align*}
    \hashf\ (lo, lm)\ k\ \lbl \eqdef{}
    & \acquire\ lo; \\
    &\DumbLet v= 
    \langkw{match}\ \mapget\ lm\ k\ \langkw{with}\\
    &\spac| \Some(b) \Ra b\\
    &\spac| \None \Ra 
    {\begin{array}[t]{l}
        \Let b = \randf\ \lbl in \\
        \mapset\ lm\ k\ b;    \\
        b
    \end{array}}\\
    &\langkw{end\ in}\\
    & \release\ lo; v
\end{align*} 
We show the presampling specification and the specification for $\hashf$ in \cref{fig:hash-spec}.
To achieve collision-freedom, we need to ensure that every value we presample to a tape generated by the
hash is different from any value previously presampled to \emph{all} tapes
generated by the hash. To be precise, suppose we have presampled a total of $s$
values on all tapes of the hash. If we want to presample a new value to a tape,
we need to pay at least $\upto{\frac{s}{\tapebound+1}}$ to sample a unique value
different from all values presampled before.  To keep track of all the values sampled
before, the interface introduces a $\hashsize$ abstract predicate that stores
the set of all values that has been presampled before. To presample onto a tape
for the collision-free hash, we need to additionally pass in a $\hashsize$
predicate to determine the amount of error needed to avoid the previous
presampled-values. The $\hashf$ specification is defined in almost the same way
as the $\lazyrandf$ specification, the view shift in the precondition performs a
case split to determine whether a key has been hashed before by looking into the
mutable map.

\begin{figure}[ht]
  \begin{subfigure}[t]{0.43\textwidth}
    \begin{align*}
      & \begin{pmatrix*}[l]
          \All  \err_O\ \prop\ \iname\ \mask\ \lbl\ \gname\ \tape\ s\ h. \\
          \spac((s+(\tapebound + 1 - s)\err_O )/(\tapebound+1)\leq \err) \wand\\
          \spac\iname\in\mask\wand\\
          \spac \ishash\ h\ \prop\ \iname\ \gname\wand\\
          \spac \hashtape\ \lbl\ \tape\ \gname \wand \\
          \spac \hashsize\ s\ \gname\wand\\
          \spac \upto{\err} \wand \\
          \spac \pupd[\mask]{}\Exists n. \upto{\err_O} \sep \hashsize\ (s+1)\ \gname\sep \\
          \spac \spac\hashtape\ \lbl\ (\tape\lapp [n])\ \gname
        \end{pmatrix*}
  \end{align*}
  \caption{Presampling Specification }
   \end{subfigure}
   ~
  \begin{subfigure}[t]{0.43\textwidth}
\begin{align*}
   & \All \iname\ \gname\ h\ \prop\ \lbl\ k\ \propB_1\ \propB_2.\\
  &\spac \hoareVH {
    \begin{array}{l}
      \ishash\ h\ \prop\ \iname\ \gname\sep\\
      (\All m. \prop\ m \wand \hashauth\ m\ \gname \wand \\
      \pupd[\top] \langkw{match}\ m!!k\ \langkw{with} \\
      \spac|\ \Some v \Ra \prop\ m \sep \hashauth\ n\ \gname \sep \propB_1\ n \\
      \spac |\ \None \Ra \Exists n\ \tape. \hashtape\ \lbl\ (n\cons\tape)\ \gname \sep  \\
      \spac \spac (\hashtape\ \lbl\ \tape\ \gname \wand \\
      \spac\spac\spac\pvs[\top] \prop\ (\lupdate{m}{k}{n}) \sep \\
      \spac\spac\spac\spac\hashauth\ (\lupdate{m}{k}{n})\ \gname\sep \propB_2\ n\ \tape)\\
      \langkw{end} 
      )
      \end{array}
  }{\hashf\ h\ k\ \lbl}{x. \propB_1\ x \lor \Exists \tape. \propB_2\ x\ \tape}[\mask]
\end{align*}
  \caption{$\hashf$ Specification }
  \end{subfigure}
  \vspace*{5mm}
  \caption{Selection of Specification of the Collision-free Hash}
  \label{fig:hash-spec}
\end{figure}

We also used this specification to derive an \emph{amortized} version of the collision-free hash, which we show in \cref{fig:amortized-hash-spec}. This hash specification has two main advantages. Firstly, clients do not need to pass a $\hashsize$ predicate as a precondition for presampling into the tape. In addition, the error credit $\err_A(\tapebound, \tapeboundB)$ to be paid  is constant as it is amortized across a fixed number of insertions $\tapeboundB$ that is decided in advance. To keep track of the maximum number of times the hash is used, clients need to give up a single $\hashtoken$ predicate; exactly $\tapeboundB$ number of these $\hashtoken$s are generated when the hash is initialized. The proof of this more complex specification is similar to that in \citet{eris} which we omit here. We emphasize that this amortized specification can be \emph{derived} from the non-amortized specification (\cref{fig:hash-spec}) without taking into account how the concurrent hash is implemented.

\begin{figure}[ht]\begin{align*}
      & \begin{pmatrix*}[l]
          \All  \err_O\ \prop\ \iname\ \mask\ \lbl\ \gname\ \tape\ h. \\
          \spac\iname\in\mask\wand\\
          \spac \ishash\ h\ \prop\ \iname\ \gname\wand\\
          \spac \hashtape\ \lbl\ \tape\ \gname \wand \\
          \spac \hashtoken\ 1\ \gname\wand\\
          \spac \upto{\err_A(\tapebound, \tapeboundB)} \wand \\
          \spac \pupd[\mask]{}\Exists n. \hashtape\ \lbl\ (\tape\lapp [n])\ \gname
        \end{pmatrix*}
  \end{align*}
  \caption{Amortized Presampling Specification}
  \label{fig:amortized-hash-spec}
\end{figure}

Just like the specification presented in \cref{sec:lazy-rand}, the specification for both the non-amortized and amortized concurrent collision-free hashes uses the probabilistic update modality in the view shift in its $\hashf$ specification to allow presampling to occur within the $\hashf$ body. As an example, consider the following program  (similar to $\lazyrace$ in \cref{sec:lazy-rand}) and its specification. 
\[\hoareV{\upto{\err_A(\tapebound, \tapeboundB)}}{
  \begin{array}{l}
    \Let h= \inithash\ \TT in \\
    \Parallel{(\hashf\ h\ 0\ (\hashalloc\ \TT))}{(\hashf\ h\ 0\ (\hashalloc\ \TT))}
    \end{array}
}{v.\Exists n. v= (n,n)}\]
Here we create an amortized hash and spawn two threads that each creates a tape and uses the tape to hash the value $0$. Because both threads are hashing the same key $0$, it should be the case that we only need to pay one constant $\err_A(\tapebound,\tapeboundB)$ for the first hash operation. However we do not know which thread is scheduled first in advance, so we cannot perform the presampling in advance before the $\hashf$ call. The probabilistic update modality allows us to perform the presampling within the $\hashf$ call in the case where the value of $m!!0$ is $\None$, indicating that this thread has been scheduled first to do the randomized sampling.

}{}

\end{document}